\documentclass[printer]{aa}

\usepackage{lscape}
\usepackage{color}
\usepackage[caption=false,font=footnotesize]{subfig}
\usepackage{graphics,graphicx}
\usepackage{rotating}
\usepackage{textcomp}
\usepackage[varg]{txfonts}
\usepackage{natbib}
\usepackage{pdflscape}
\usepackage{longtable}
\usepackage{multicol}
\usepackage{supertabular}

\begin{document}


\title{Measurements of Forbush decreases at Mars: \\ both by MSL on ground and by MAVEN in orbit}


\author{
Jingnan Guo\inst{1}
\and Robert Lillis \inst{2}
\and Robert~F. Wimmer-Schweingruber\inst{1} 
\and Cary Zeitlin \inst{3}
\and Patrick Simonson\inst{1,4}
\and Ali Rahmati \inst{2}
\and Arik Posner\inst{5}
\and Athanasios Papaioannou\inst{6}
\and Niklas Lundt\inst{1}
\and Christina O. Lee\inst{2}
\and Davin Larson \inst{2}
\and Jasper Halekas\inst{8}
\and Donald~M. Hassler\inst{7}
\and Bent Ehresmann\inst{7}
\and Patrick Dunn \inst{2}
\and Stephan B\"ottcher\inst{1}
}

\institute{ {Institute of Experimental and Applied Physics, Christian-Albrechts-University, Kiel, Germany. \email{guo@physik.uni-kiel.de}}
\and {Space Sciences Laboratory, University of California, Berkeley, USA}
\and {Leidos, Houston, Texas, USA}
\and {Physics and Astronomy Department, College of Charleston, South Carolina, USA}
\and {NASA Headquarters, Science Mission Directorate, Washington DC, USA}
\and {Institute for Astronomy, Astrophysics, Space Applications and Remote Sensing, National Observatory of Athens, Greece }
\and {Southwest Research Institute, Boulder, CO, USA}
\and {Department of Physics and Astronomy, University of Iowa, Iowa City, Iowa, USA}
}

\abstract{
The Radiation Assessment Detector (RAD), on board Mars Science Laboratory's (MSL) Curiosity rover, has been measuring ground level particle fluxes along with the radiation dose rate at the surface of Mars since August 2012. 
Similar to neutron monitors at Earth, RAD sees many Forbush decreases (FDs) in the galactic cosmic ray (GCR) induced surface fluxes and dose rates. These FDs are associated with coronal mass ejections (CMEs) and/or stream/corotating interaction regions (SIRs/CIRs). 
Orbiting above the Martian atmosphere, the Mars Atmosphere and Volatile EvolutioN (MAVEN) spacecraft has also been monitoring space weather conditions at Mars since September 2014. The penetrating particle flux channels in the Solar Energetic Particle (SEP) instrument onboard MAVEN can also be employed to detect FDs.
For the first time, we study the statistics and properties of a list of FDs observed in-situ at Mars, seen both on the surface by MSL/RAD and in orbit detected by the MAVEN/SEP instrument. Such a list of FDs can be used for studying interplanetary CME (ICME) {propagation} and SIR {evolution} through the inner heliosphere. The magnitudes of different FDs can be well-fitted by a power-law distribution. 
The systematic difference between the magnitudes of the FDs within and outside the Martian atmosphere may be mostly attributed to the energy-dependent modulation of the GCR particles by {both} the pass-by ICMEs/SIRs {and} the Martian atmosphere. }


\titlerunning{Forbush decreases at Mars}
\authorrunning{GUO ET AL.}
\maketitle 

\section{Introduction and Motivation}\label{sec:intro}
Galactic cosmic rays (GCRs) are omnipresent in the heliosphere, and their intensity varies as a result of modulation in the heliosphere by the change of solar magnetic activity.
Forbush decreases (FDs) are identified as a temporary and rapid depression in the GCR intensity, followed by a comparatively slower recovery phase and typically last for a few days \citep{lockwood1971forbush, belov2008forbush, cane2000}. FDs were first discovered by \citet{forbush1938} using ground-based {measurements} at Earth and have since been frequently reported and studied in many publications \citep[see, e.g., the review article of][]{cane2000}. 
Generally, FDs are caused by interplanetary disturbances related to solar coronal mass ejections (CMEs) and/or the streaming interaction regions (SIRs) as well as corotating interaction regions (CIRs) which are caused by the interaction of fast solar wind streams which catch up with slow wind streams. 
Thus, interplanetary disturbances that give rise to FDs can be either sporadic or of recurrent nature \citep{belov2008forbush}.

The majority of FDs are of sporadic character and caused by interplanetary CMEs (ICMEs) which are often associated with a shock front followed by an ejecta both modulating the intensity of GCRs in the interplanetary space. 
Such decreases result {from} the interplay of two processes. In particular, once GCRs encounter the ICME an almost linear decrease in the GCR flux is observed during their passage from the ICME turbulent sheath region \citep[the region between the ICME driven shock and the starting of the ejecta; ][]{barnden1973large, wibberenz1998transient}. Furthermore, the magnetic configuration of ICMEs can lead to a subsequent local decrease of the GCR intensity, once the ICME moves over the observer \citep{richardson2011}.

Alternatively, the recurrent FDs are related to the interaction of high-speed streams (HSS) with background solar wind and they often occur periodically as the Sun rotates. 
The HSS-related FDs at Earth on average have smaller magnitudes than ICME-related ones \citep{belov2014}. 
For a comprehensive review the reader is referred to \citet{richardson2004energetic}.

At Earth, ground level detectors such as neutron monitors, muon telescopes \citep[e.g.,][]{simpson1983}, or even water-Cherenkov detectors  \citep{dasso2012scaler} 
with different geomagnetic cutoff rigidities (rigidity is particle momentum per unit charge) have been the main source for studying FDs and the related space weather conditions for more than half a century. 
It has been found that there is an energy dependence on the FD amplitude \citep{cane2000, lingri2016solar} and its recovery time \citep{usoskin2008forbush} due to the energy-dependent modulation of the GCR: the strength of the modulation, the amplitude of the FDs, and the recovery time are anti-correlated with the kinetic energy of the GCR particles. 
{Moreover, due to different velocities and highly complex structures of the ICME encountered, it has been shown that the resulted FDs exhibit a large variation in their amplitudes even for the same energies of GCRs \citep{belov2015galactic}.} 
At the same time, the intensity variations of GCRs at Earth during FDs have provided an alternative tool for the identification of the arrival of CME-driven shocks \citep{papailiou2012precursor} and have also provided a reconstruction of the interplanetary conditions even in the absence of in-situ plasma measurements \citep{papaioannou2010analysis}. 
The study of FDs is still a very dynamic research field. Their "standard picture" proposed by \citet{barnden1973large} and \citet{cane2000} was questioned by \citet{jordan2011revisiting}. Recently, a new classification has been proposed by \citet{raghav2017forbush} who argue that localized structures within the shock sheath and magnetic cloud may also have a significant role in influencing the FD profile.

Due to the diversity of solar sources and their dynamic interactions throughout the heliosphere, the properties of FDs are very diverse: they can be large or small, short-term or long lasting, with fast or gradual decrease, with full recovery or without it at all, falling with two steps or not, with simple or complicated time profile and so on. 
Another reason for the diversity of FD properties is that the observation of an FD event is usually studied at one point in the interplanetary space, mostly on and near Earth, while the same FD may look different at other locations in the heliosphere. 
This is related to the fact that an ICME's intensity, speed, geometry and interaction with the ambient solar wind may change drastically as it propagates outwards from the Sun through the heliosphere. 
On the other hand, HSS often evolve dynamically as they continuously interact with the background plasma and even with the propagating ICMEs. CIR or SIR related shocks are generally developed beyond 1AU.   
Therefore it is very important to study FD properties at other heliospheric locations away from Earth and at solar distances other than 1 astronomical unit (au). 
Due to the limited number of spacecraft in the heliosphere away from Earth, very few studies have addressed GCR modulation by ICMEs at other heliospheric distances \citep[e.g.,][]{witasse2017interplanetary, guo2015modeling}. 

In this paper and for the first time, we identify and study a list of FDs measured at planet Mars ($\sim$ 1.5 au from the Sun), both on the surface and outside the Martian atmosphere. 
We use two sets of observations: (a) dose rate from energetic particles on the ground at the location of Gale Crater measured by the Radiation Assessment Detector \citep[RAD,][]{posner2005, hassler2012} on board Mars Science Laboratory's (MSL) Curiosity rover \citep{grotzinger2012mars} and (b) particle count rates collected by the solar energetic particle \citep[SEP,][]{larson2015maven} instrument aboard the Mars Atmosphere and Volatile EvolutioN \citep[MAVEN,][]{jakosky2015initial} spacecraft orbiting Mars.
As a first step of investigation, we do not specify the interplanetary disturbance cause of the FD, i.e., whether it is ICME or HSS related. 
Rather, we select all identifiable FD events seen by both MSL/RAD and MAVEN/SEP over a time period of one Martian year (from 2014-10-01 to 2016-09-12) and study, comparatively, the statistics of FD magnitudes (i.e., drop percentage) and the correlation between the FD magnitudes on the surface of Mars seen by MSL/RAD and those in orbit detected by MAVEN/SEP.
These first results give us (i) an overall view of FD properties at Mars, (ii) how they differ from those at Earth and (iii) how the Martian atmosphere potentially modifies the amplitudes of FDs.
In section \ref{sec:space_weather_mars}, we present an example of space weather scenarios observed at Mars during a period when both MAVEN and MSL measurements are present; in section \ref{sec:measurements_GCR}, we describe in detail the two instruments used for detecting GCRs and subsequent FDs (RAD on the surface and SEP in orbit around Mars); 
in section \ref{sec:FDatMars}, we select, analyze and compare the FDs observed during these two years both on the surface and in orbit of Mars; and finally in section \ref{sec:discussion}, we summarize the results and discuss the general FD properties at Mars, the Martian atmospheric influence on the FD amplitude and the possible causes of differences of FDs at Earth compared to those at Mars.

\section{Space weather scenarios at Mars}\label{sec:space_weather_mars}

FDs at Earth and their association with HSS, shocks and ICMEs have been studied extensively by previous authors using both in-situ solar wind and IMF measurements as well as ground-based monitors \citep[e.g.,][]{cane2000, belov2008forbush, belov2014, masias2016superposed}.
FDs at Mars and how their properties are related to different solar wind and plasma conditions are a relatively new topic in the field of space weather due to limited in-situ observations at Mars. 
The recent measurements by both MSL/RAD and MAVEN have provided a great opportunity for this study.  
Here we briefly show an example of various space weather conditions at Mars and how they are related to the occurrence of FDs seen at Mars.

\subsection{Data used in the analysis}
\begin{figure*}[pht]
	\centerline{\includegraphics[width=0.80\textwidth]{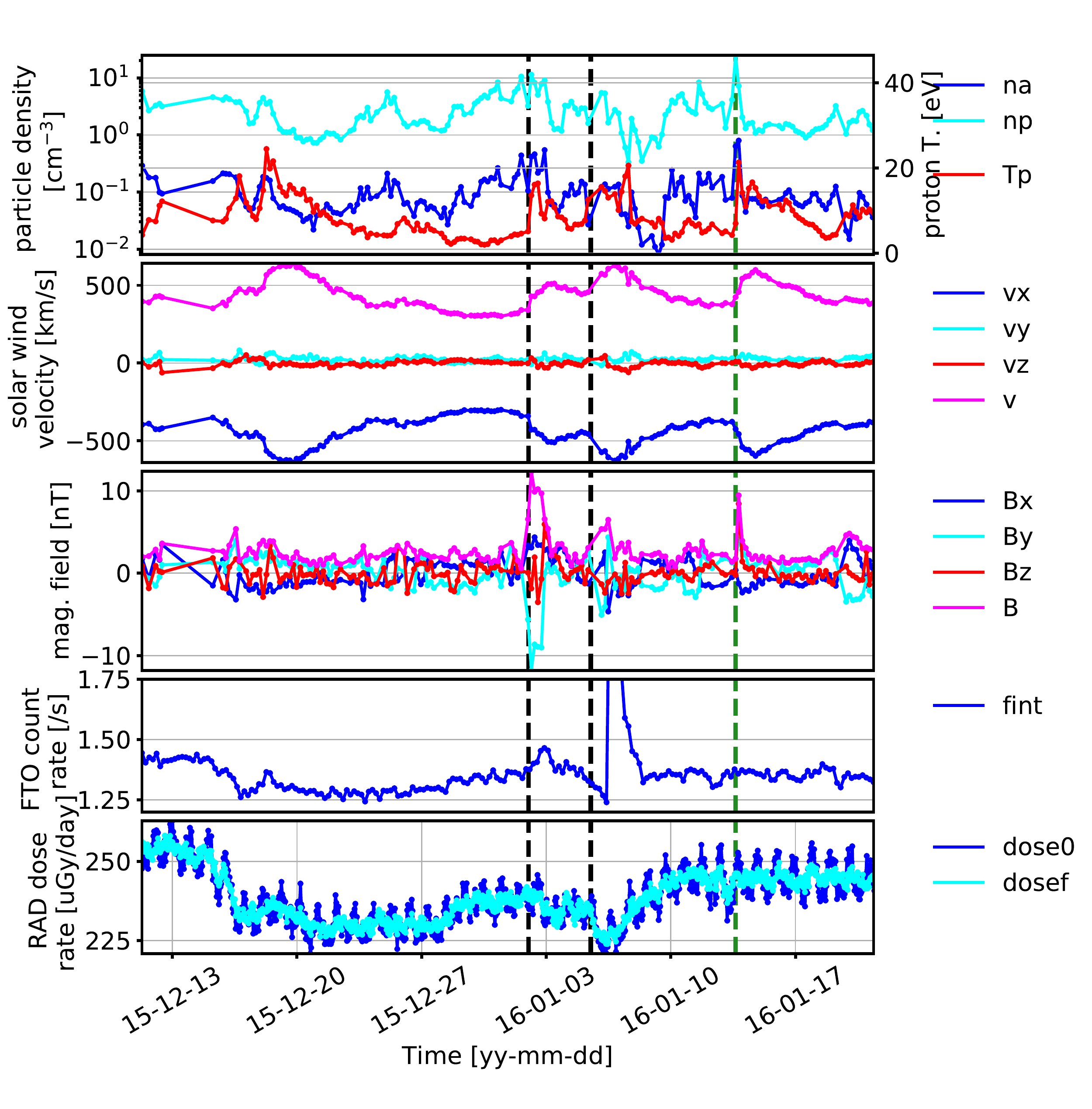}}
	\caption{MAVEN and MSL/RAD measurement of several possible HSS, CME, FD and SEP events at Mars from 2015-12-11 until 2016-01-21. 
		The top two panels show the parameters measured by MAVEN/SWIA in the upstream solar wind region: proton temperature ($T_p$), particle density of alphas ($n_a$) and protons ($n_p$), solar wind velocity ($v_x$, $v_y$ and $v_z$) and its speed ($v$). The third panel shows the MAVEN/MAG measurement of IMF ($B_x$, $B_y$ and $B_z$) and its strength ($B$) in the upstream solar wind region. All vectors are in MSO coordinates.
		The next panel shows the integrated count rate 'fint' approximately corresponding to protons $\ge 100$ MeV in the {MAVEN/SEP sensor 1 (averaged values of A and B sides)}. 
		The last panel shows the dose rate measured by MSL/RAD on Martian surface with 'dose0' standing for the original dose rate data while 'dosef' for the filtered data using a notch filter to remove the variations at one sol frequency. Vertical lines denote the identified shocks with black and green lines for ICME- and SIR- related shocks respectively.}\label{fig:maven-msl-CME}
\end{figure*}

To show an example of space weather conditions at Mars, we use the MSL/RAD and MAVEN/SEP measurements together with data from another two MAVEN instruments: the Solar Wind Ion Analyzer \citep[SWIA,][]{halekas2015solar} measuring the solar wind plasma parameters and the Magnetometer \citep[MAG,][]{connerney2015maven} carrying out vector measurements of the interplanetary magnetic field (IMF).
We select a period from 2015-Dec-11 until 2016-Jan-21 during which MSL/RAD saw several major FD events on the surface of Mars. 
Figure \ref{fig:maven-msl-CME} shows the MAVEN/SWIA, MAVEN/MAG, MAVEN/SEP 
and MSL/RAD measurements during this period including, from top to bottom, the proton temperature ($T_p$), particle density of alphas ($n_a$) and protons ($n_p$), solar wind velocity ($v_x$, $v_y$ and $v_z$ in Mars-Solar-Orbital (MSO) coordinates) and its speed ($v$), IMF vector components ($B_x$, $B_y$ and $B_z$ in MSO) and its strength ($B$), the integrated FTO (three detectors abbreviated as F, T and O) count rate mainly corresponding to protons $> 100$ MeV through {the MAVEN/SEP sensors (more detail in Section \ref{sec:MAVEN/SEP} and also in \citet{larson2015maven})}, and finally the dose rate measured by MSL/RAD on the Martian surface (more detail in Section \ref{sec:MSL/RAD} and also shown in Figure \ref{fig:maven-msl-all}).
Corrections following data processing methods shown in \citet{connerney2015first} have been applied to remove artifacts related to solar array circuit excitation in the MAG data. We also note that the SWIA and MAG data has been selected during each MAVEN orbit taking into account the Martian bow shock structure and the solar wind interactions with Mars \citep{halekas2017structure}. 
For our purposes here, only the upstream data has been selected. 
As the Martian bow shock has significant variability in location, it is possible to get interspersed periods with and without solar wind coverage. 
\subsection{An example of Martian space weather scenario}
As shown in Figure \ref{fig:maven-msl-CME}, there were several significant FDs with a complex structure and evolution and a magnitude (biggest variation normalized to the value before the onset of the FD) of up to $\sim 10\%$ occurring from around 2015-12-14 lasting for nearly a month as seen by MSL/RAD and also in the MAVEN/SEP integrated FTO count rate.  
The December FD correlated with an extended HSS structure indicated by the significantly enhanced solar wind velocity up to $\sim 40\%$. As the MAVEN orbit did not enter the undisturbed solar wind for some days resulting in a gap in MAVEN/SWIA and MAVEN/MAG upstream data used here, we couldn't clearly identify a shock arrival at Mars at the onset of the FD event. 
With a small recovery from 2015-12-17 to 2015-12-18, the FD took its second step reaching its nadir around 2015-12-21 and lasted for 6-7 days. 
The WSA-ENLIL+Cone model simulations (https://iswa.gsfc.nasa.gov/IswaSystemWebApp/; {standard runs from the Mars mission support interface are utilized here and throughout the paper}) show that at this period, there were two HSS structures passing by Mars. So it is likely that the second step of the FD was related to the second structure before the full-recovery of the first FD. 
As the solar wind speed slowly recovered to its normal values at around 2015-12-25, there seemed to be a few days of delay of the recovery of the FD. 

On 2016-01-02, there was an ICME-related shock arriving at Mars with a clear enhancement in the magnetic field data ($\sim 4$ times). This shock arrival is plotted in Figure \ref{fig:maven-msl-CME} as the first dashed vertical line and has been identified by \citet{lee2017maven} as related to an M1.8 flare in the NOAA active region (AR) 12473. 
The integrated count rate of the MAVEN FTO channels (explained in Section \ref{sec:MAVEN/SEP}) sees a small enhancement related to the shock accelerated particles. 
As most of the these particles can not penetrate through the Martian atmosphere, on the surface RAD sees an FD lasting about two days with a small recovery. 
Around 2016-01-03, it is possible to identify an ICME ejecta passing Mars via the rotation of the $B_x$ and $B_y$ component and the decrease of the proton temperature $T_p$ as shown in the MAG and SWIA measurements. 
Also visible is a small second-step decrease between 2016-01-03 and 2016-01-05 during the pass-by of the magnetic ejecta. 

On 2016-01-06, there was another clear shock arrival at Mars as shown in Figure \ref{fig:maven-msl-CME} with the middle dashed vertical line. This has been also reported by \citet{lee2017maven} as related to an ICME launched from the backside of the Sun (with respect to Earth) while the WSA-ENLIL+Cone modeled results do not show this ICME passing by Mars. 
The MAVEN integrated FTO count rate showed a significant and sudden enhancement lasting from 2016-01-05 to 2016-01-08 due to the shock accelerated ions and electrons. 
On the other hand, the FD seen by RAD lasted for about 6 days and finally fully recovered on 2016-01-11.
 
Only two days later on 2016-01-13 (denoted by the third vertical dashed line), there was another shock arrival clearly seen in the IMF and solar wind velocity. This shock has also been reported by \citet{lee2017maven} as related to an SIR. Meanwhile, both RAD dose and SEP/FTO observations showed a small but visible FD. 

The above-described dynamic and complex space weather scenarios at Mars are in fact not uncommon and the occurrence of the FDs can be due to shocks, CMEs, HSS-related SIRs and CIRs or even a combination of these effects. A recent overview of space weather at Mars measured by MAVEN from Nov 2014 until June 2016 contained a list including most of the major heliospheric events seen at Mars during this period \citep{lee2017maven}. 
In the current paper, we focus on the statistics and magnitudes of FDs seen by both MAVEN/SEP/FTO and MSL/RAD on the surface of Mars over a period of 2 Earth years. 
Since we aim to have good statistics and study the atmospheric effects on the FDs, we hereby do not yet differentiate the interplanetary disturbance cause of the FD, i.e., whether it is ICME or HSS related. 
Therefore in the following, we will only make use of the MAVEN/SEP and MSL/RAD measurements to identify FDs and their magnitudes both in orbit and on the surface of Mars.

\section{Measurements of GCRs at Mars}\label{sec:measurements_GCR}
\subsection{GCR measurement by MAVEN/SEP}\label{sec:MAVEN/SEP}
\begin{figure*}
\centerline{\includegraphics[width=0.95\textwidth]{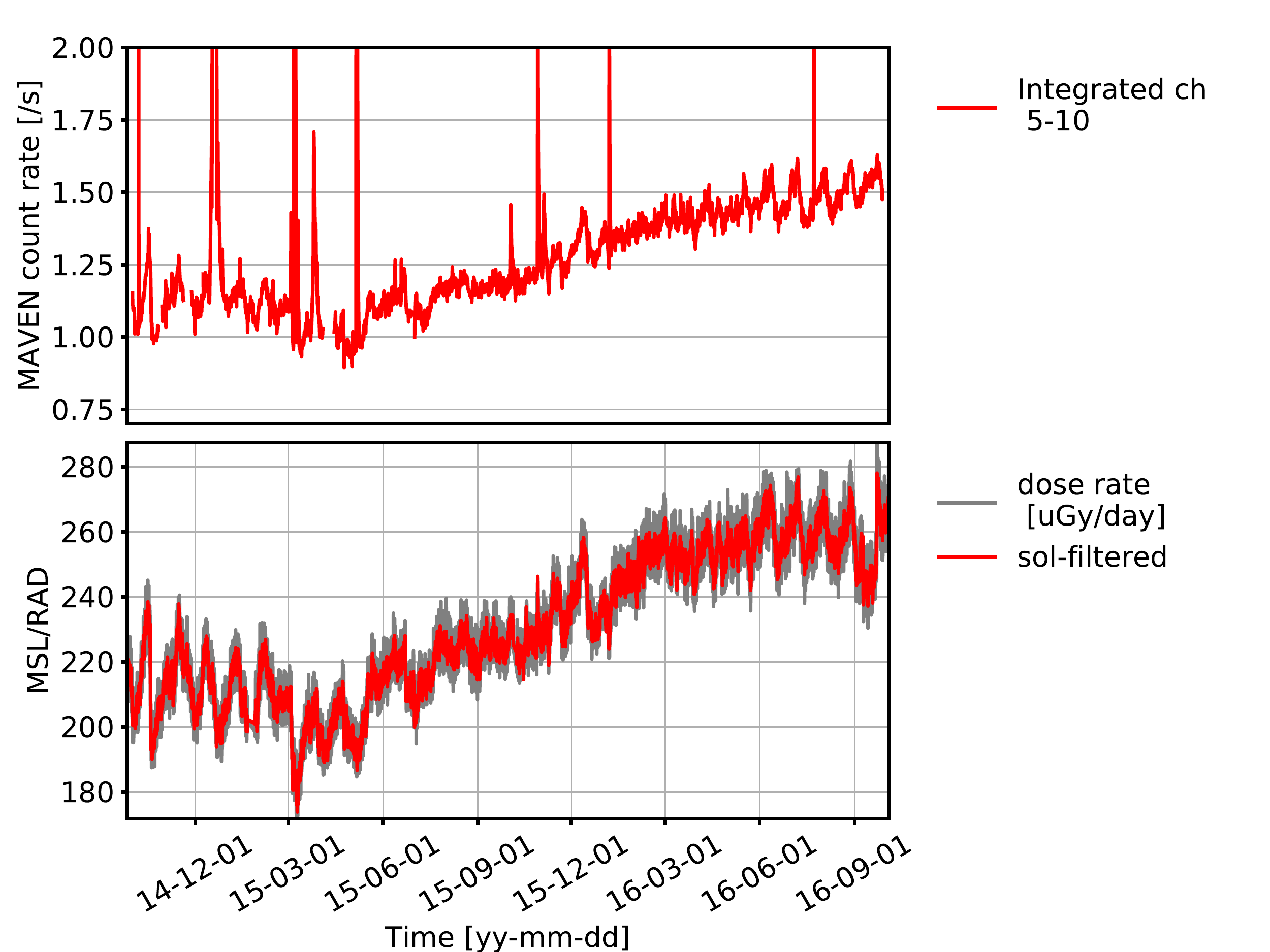}}\caption{\textit{Top:} integrated flux rate corresponding to protons larger than 100 MeV (FTO channels 5 to 10 of the MAVEN/SEP instrument are integrated). \textit{Bottom:} plastic dose rate measured by MSL/RAD at Gale Crater with the gray color for the original data while red for the filtered data (removing the variations at one sol frequency). The time range of the data is from 2014-Sep-30 until 2016-Sep-30.}\label{fig:maven-msl-all}
\end{figure*}



MAVEN has been continuously monitoring the local space weather conditions around Mars since October 2014 following its arrival at the red planet. 
The SEP instrument on board was designed to measure energetic solar particles and pickup ions at Mars with an energy range of 20 to 6000 keV for ions and 20 to 1000 keV for electrons as shown in Fig. 6 in \citet{larson2015maven}. 
These are however not the suitable energy ranges for identifying FDs, which are depressions in the highly energetic (MeV/nuc - GeV/nuc) GCR fluxes. 
Alternatively, we exploit particles penetrating through the whole SEP detector set and we try to select protons with energies larger than 100 MeV which in most cases are GCRs during solar quiet time. 
To do so, we can estimate the energy range of detected particles by utilizing the correlation between incident energy and deposited energy of protons penetrating through all three FTO detectors \citep{bethe1953passage}.
Specifically speaking, penetrating protons (through all three FTO detectors) with energies between 100 MeV and 200 MeV may lose between 0.79 and 1.21 MeV in the detectors (corresponding to MAVEN SEP/FTO channels 9 and 10); penetrating protons with energies between 800 MeV and 1 GeV may lose between 0.25 and 0.36 MeV in the detectors (corresponding to channels 5 and 6). 
For penetrating electrons, it is however much more difficult to identify the energy range of their incident energies \citep{larson2015maven} and if electrons with energies $\ge$ 600 keV are abundant, {they may also be recorded in FTO channels 5 to 10. 
Thus one may not be able to clearly discriminate between counts from such electrons versus penetrating protons in the MAVEN/SEP FTO detected energy channels.
Nevertheless, as GCRs are composed of mainly protons, it is reasonable to assume that FTO penetrating channels measure mainly GCR protons during solar quiet times.}
The integrated count rates from channels with deposited energies smaller than 1.21 MeV and incident proton energies larger than 100 MeV, i.e., the summed fluxes from channels 5 to 10, are employed in the current study. 

Since MAVEN orbits around Mars with a highly elliptical orbit, the planet itself may shadow a significant amount of GCR fluxes when MAVEN is very close to Mars \citep{lillis2016shadowing, luhmann2007solar} resulting in a periodic decrease in the integrated count rate. As particles with such high energies are not affected by the weak Martian magnetosphere, the shadowing effect of integrated flux only depends on the orbiting altitude, which is periodic with each orbit. Therefore we average the data in each MAVEN orbit (with a duration of 4.5 hours) to remove the periodic oscillation in the data. The time resolution is thus reduced from 8 min to 4.5 hours which is still adequate for identifying FD effects, which typically have durations lasting for days.

Such orbit-averaged flux rate of channels 5-10 integrated together, corresponding to protons with energies larger than 100 MeV (with some contributions of $\ge$ 600 keV electrons as mentioned earlier and also explained in detail in the appendix) is plotted in Fig. \ref{fig:maven-msl-all}.
Data of about two years duration from 2014-Sep-30 until 2016-Sep-30 are plotted here. 

\subsection{GCR measurement by MSL/RAD}\label{sec:MSL/RAD}

MSL/RAD is an energetic particle detector and it has been carrying out radiation measurements of both particle types and fluxes together with dose and dose equivalent rates during its 11-month cruise phase \citep{zeitlin2013, posner2013, guo2015cruise, ehresmann2016charged} and later on the surface of Mars since the landing of MSL in August 2012 \citep{hassler2014, ehresmann2014, rafkin2014, koehler2014, wimmer2015, guo2015modeling}.
On the surface of Mars, RAD measures a mix of primary GCRs or SEPs and secondary particles generated in the atmosphere including both charged \citep{ehresmann2014} and neutral \citep{koehler2014, guo2017neutron} particles.
RAD measures dose in two detectors: the silicon detector B and the plastic scintillator E \footnote{Radiation dose rate is a key quantity used to evaluate the energetic particle environment. Dose is detected as the energy deposited by particles per unit detector mass with a unit of J/kg (or Gy). The plastic scintillator E has a composition similar to that of human tissue and is also more sensitive to neutrons than silicon detectors.}. 
Due to its much bigger geometric factor, the dose rate measured in detector E has much better statistics and is a very good proxy for GCR fluxes. 

In the lower panel of Fig. \ref{fig:maven-msl-all}, we present the MSL/RAD dose rate measured at Gale Crater on the surface of Mars during the same period of the MAVEN data from 2014-Sep-30 until 2016-Sep-30. It is clearly shown that the MAVEN high-energy flux rate and the RAD dose rate are very well correlated in the long-term evolution. 
After excluding the SEP events present in the MAVEN data, the cross-correlation coefficient between the two data sets has been calculated to be about 0.87. 

Due to the weakened solar modulation of GCRs as the current solar cycle declines, the average dose rate in RAD has increased by about 25\% in the period investigated here and the average MAVEN integrated count rate has increased by about 38\%. 
Note that this increase ratio is stronger at MAVEN orbiting Mars compared to at MSL on the surface of Mars. 
This is possibly related to the fact that the Martian atmosphere shields out especially the less-energetic incident particles which are more effectively modulated by solar activity \citep{guo2017dependence}. More detailed investigations and explanations of the Martian atmosphere affecting the modulation of GCRs will be discussed later in Section \ref{sec:discussion}. 

Analysis using RAD measured count rates has also been carried out and the data also clearly show the long-term GCR flux variation as the solar activity evolves. However, for short-term FD signals, the dose rate data appear to be a better proxy. This is because the count rate data generally (1) contain more noise due to different instrumental cuts and geometric restrictions and (2) include a large amount of secondaries such as electrons, positrons, muons, gammas, neutrons, and so on which are generated in the atmosphere and more susceptible to atmospheric disturbances. 
Alternatively the contribution by such secondaries to the dose data is relatively small while the contribution by primary and heavy-mass GCRs to dose rate is larger than to the count rate. 
This is supported by the clear anti-correlation between RAD measured dose rate and the surface pressure, suggesting a shielding effect of the atmospheric depth against the GCR-induced dose \citep{rafkin2014}.

\subsection{Filtering out diurnal variations in MSL/RAD}\label{sec:notch_filter}
The Martian atmosphere exhibits a strong thermal tide excited by direct solar heating of the atmosphere on the dayside and strong infrared cooling on the nightside. Heating causes an inflation of the atmosphere with a simultaneous drop in surface pressure and column mass.
At Gale Crater, the thermal tide produces a daily variation in column mass of about $\pm 5\%$ relative to the median, as measured by the Rover Environmental Monitoring Station (REMS) \citep{haberle2014preliminary}. 
This diurnal change of the atmospheric depth causes daily oscillations of the dose rate measured by RAD of up to $\sim$ 2.5\%: when the pressure (column mass) increases during the night, the total dose rate decreases; when the pressure decreases during the mid-day, the total dose rate increases \citep{rafkin2014}. 
We also note that the Martian global CO$_2$ condensation cycle may cause up to 20\% of the seasonal variation of the atmospheric depth which would result in bigger variations in the RAD measured dose rate \citep{guo2015modeling}. However for identification of  short term FDs and their properties, this long-term influence can be ignored. 
The Martian atmospheric shielding effect on the accumulated dose rate has also been studied using both modeled and measured data \citep{guo2017dependence}. It was shown that the atmospheric shielding of the GCR dose rate depends on both solar modulation and atmospheric conditions; and this dose-pressure correlation is a non-linear effect.

For identifying CME or CIR related Forbush decreases in the dose rate, which mostly have magnitudes on the order of a few percent (compared to $\sim$ 2.5\% of the diurnal oscillations due to pressure disturbances), it is important to account for pressure-induced diurnal variations in the data. 
Simply averaging the data into daily values would indeed remove the daily variations; however, the one-sol time resolution may become insufficient for better defining onset times.
In order to remove the diurnal oscillations while keeping an adequate time resolution ($\sim$ 32 minutes), a digital filter is applied to the data. Since the frequency of the diurnal disturbance is known and constant, a notch filter tuned to remove all the multiples of 1 sol harmonics can be used \citep{parks1987digital}.
To generate the filtered data $f(t)$, the convolution of the original data $d(t)$ (where t is time and d is the measurement) with a filter function $h(t)$ can be converted into a product $f(s)=d(s)\cdot H(s)$ in the Laplace-transformed domain. 
As the data collection is not continuous, a bilinear transformation shown in Equation \ref{eq:bilinear} is used for obtaining the time-discretized transfer function $H(z)$ from the continuous time transfer function $H(s)$ \citep{parks1987digital}:
\begin{equation}
\label{eq:bilinear}
s=\frac{2}{T_s}\frac{z-1}{z+1},
\end{equation}
where $s$ is the Laplace transform variable; $z$ is the Z-transform variable and $T_s$ is the sampling time (which is a constant of 1/44 sol; about 44 or 88 observations are collected per sol depending on the operational mode). For periods when observations are not recorded at such a frequency, we bin the data into 1/44 sol of sampling time. 

The notch filter's transfer function in the Laplace domain is shown in Equation \ref{eq:notch} where $\omega_0=\frac{2\pi}{1sol}$ is the fundamental pulsation of the disturbance; $N$ is the maximum harmonic order that should be eliminated (12 is an optimized chose here as constrained by the sampling frequency); and $\delta$ is the bandwidth of the filter which is set to be 0.1 being sufficiently smaller than 1 sol.
\begin{equation}
\label{eq:notch}
H_{notch}(s, N)=\sum_i^N\frac{s^2+\left(i\omega_0\right)^2}{s^2+2\delta i \omega_0s+\left(i\omega_0\right)^2}
\end{equation}

Considering the sampling frequency (44 per sol) is not very high compared to the frequencies that should be filtered out, frequency pre-warping should be used before the discretization process. This implies that the continuous-time filter should be applied as following:
\begin{equation}
\label{eq:notch_prewarp}
H^{pre}_{notch}(s, N)=\sum_i^N\frac{s^2+\left( \frac{2}{T_s}\tan(i\omega_0 \frac{T_s}{2})\right)^2}{s^2+2\delta \frac{2}{T_s}\tan\left(i\omega_0 \frac{T_s}{2}\right)s+\left(\frac{2}{T_s}\tan\left(i\omega_0 \frac{T_s}{2}\right)\right)^2}
\end{equation}

The filtered dose rate data without diurnal oscillations compared to the original data is shown in the bottom panels of Figure \ref{fig:maven-msl-CME} and \ref{fig:maven-msl-all}. It is readily seen that the onsets and magnitudes of FDs are much more clearly seen in such processed data.

\section{FD events at Mars measured by both MSL and MAVEN}\label{sec:FDatMars}

\subsection{Selection of FDs in the data}\label{sec:selectionFDs}
\begin{table*}
\begin{center}
\caption{The uncertainties of the manual identification procedure for the onset time, nadir time, drop duration and drop ratios at MSL and at MAVEN. \label{table:uncertainty}}
\begin{tabular}{rrrrr}
\hline
 $\delta_{onset}$ (day) & $\delta_{nadir}$ (day) & $\delta_{drop\ duration}$ (day)& $\delta_{drop\ ratio\ MSL}$ (\%) &$\delta_{drop\ ratio\ MAVEN}$ (\%) \\
0.68 &  0.55 & 1.05 & 1.38 &  1.94 \\ 
\hline
\end{tabular}
\end{center}
\end{table*}

The MSL/RAD dose rate and the MAVEN/SEP count rate over the period of about two years shown in Figure \ref{fig:maven-msl-all} are then used as the database to select FDs at Mars. {Although the MAVEN/SEP FTO high-energy particle measurements are a good proxy for detecting GCRs}, there are advantages to using MSL/RAD for detecting FDs due to (a) MSL/RAD on the Martian surface measures almost exclusively GCRs and their secondaries during solar quiet times and (b) the dose data has better statistics due to a larger geometric factor of the RAD plastic detector. 

The identification and selection of FDs has always been a procedure that is difficult to realize via automatic routines. The most reliable FD recognitions have been mainly carried out manually \citep[e.g.,][]{belov2008forbush} or half automatically with a post-process of manual selection (private communication with Athanasios Papaioannou).
To select a series of FDs, we first go through the RAD dose rate data and manually define a list of FDs together with the time and dose values of the onset (the time before the decrease) and the nadir point (lowest dose rate value) for each event.  
We then scan through the MAVEN/SEP/FTO integrated count rates ($\geq 100$ MeV) to check if an FD is present in the data during the same time period; if so, we also note down the count rates of different FTO channels at the FD onset and nadir. 
For each event we calculate the drop ratio which is the magnitude of the drop (difference of the values at the onset and the nadir point) divided by the onset value at MSL and MAVEN respectively.  

In order to (a) estimate the uncertainties of the manual identification of FDs and (b) generate an FD list as complete as possible, a second person repeated the above procedure independently to generate a second list. The first list contains 98 FDs in the MSL data, among which 68 events are also seen as FDs at MAVEN. 
The second list contains 101 FDs in the MSL data, among which 54 events are also seen as FDs at MAVEN. 
For each of the two independent lists, a good portion of the events selected at MSL are not identified at MAVEN. There are three main causes for this: (a) there are occasionally observational gaps by MAVEN; (b) some events can not be clearly defined as an FD from the MAVEN/SEP measurements due to lower statistics and higher fluctuations in the data; or (c) quite often an FD is detected as an increase at MAVEN, possibly due to enhanced particle fluxes accelerated by shocks associated with the ICME, SPE or SIRs. 
As stated earlier, counts triggered from $\ge$ 600 keV electrons are not distinguished from those triggered  $\ge$ 100 MeV protons in the SEP/FTO detected energy channels. However, as shown in the Appendix of this study, most of the enhanced particle fluxes measured in FTO channels 5-10 are predominantly due to SEP electrons. 

To estimate the reliability of the manual identification procedure, we compared the two independent lists and found 80 matching events, i.e., the drop durations of two FDs from each independent list are overlapping and the two selections are in fact the same event. 
For each matching pair, we calculate the difference of the onset time, nadir time, drop duration, drop ratios at MSL and at MAVEN. The average of these differences over 80 matching pairs are then considered as the uncertainty of our manual identification procedure and they are given in Table \ref{table:uncertainty}. These uncertainty values are further propagated through the statistical analysis of these data as shown in the following section.

We have also merged the two lists and included the common events as well as other events that are present only in one of the two lists. Such a merged list contains 121 FD events seen by MSL in which 77 events are also detected by MAVEN. 
These 121 events with their onset, duration and drop ratio information are shown in the table in the appendix of this paper. 
We note that the uncertainties of the time and drop ratio in Table \ref{table:uncertainty} should be applied when quantitatively utilizing and analyzing the FD data list.  

\subsection{Statistics of FDs at Mars} \label{sec:satisticsFDs}

\begin{figure*}[tb!]
\begin{tabular}{cc}
\subfloat[FD histogram of MSL FDs] {\includegraphics[trim=5 0 20 20,clip, scale=0.45] {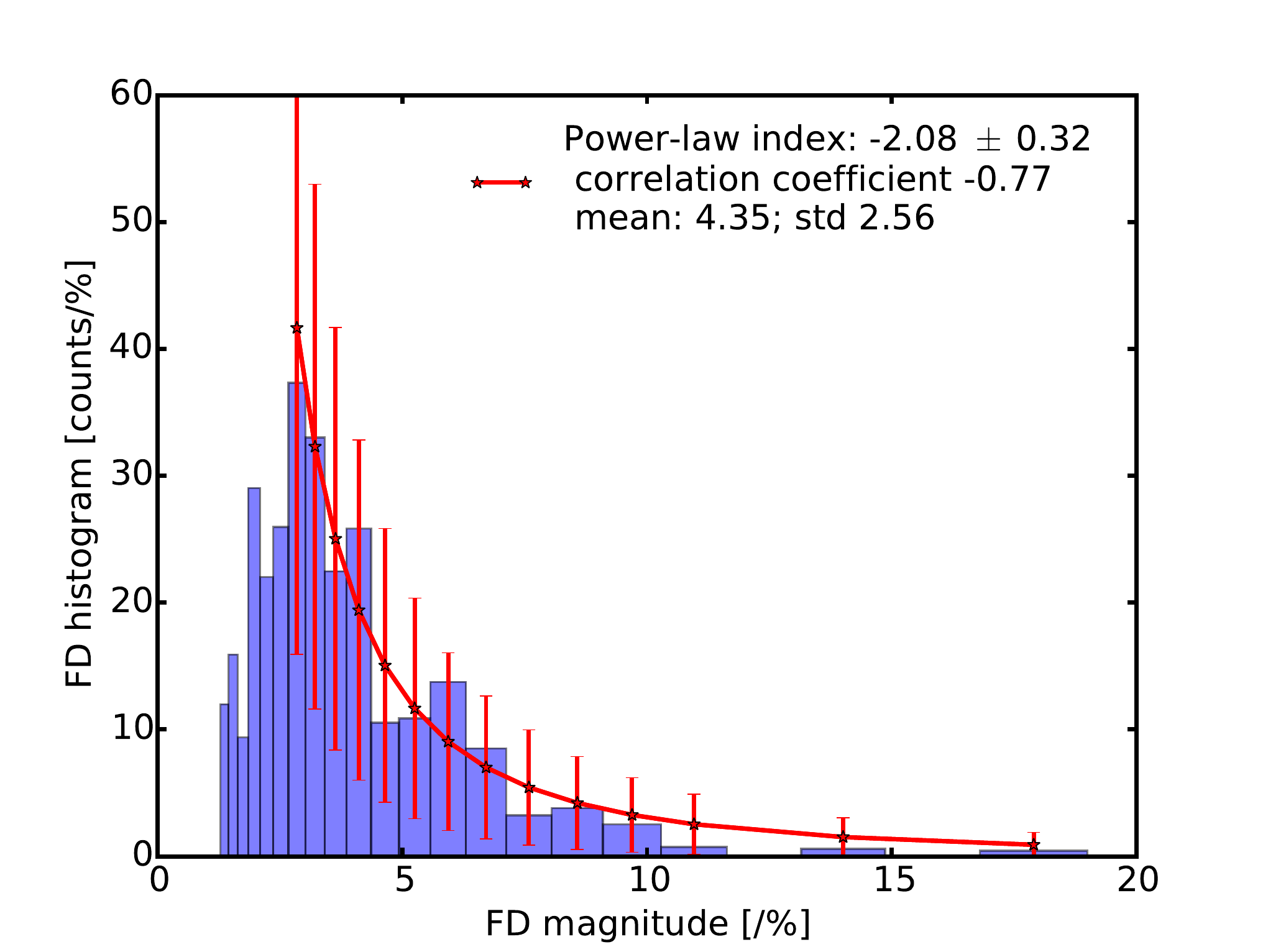}} &
\subfloat[FD histogram of all MAVEN FDs] {\includegraphics[trim=5 0 20 20,clip, scale=0.45] {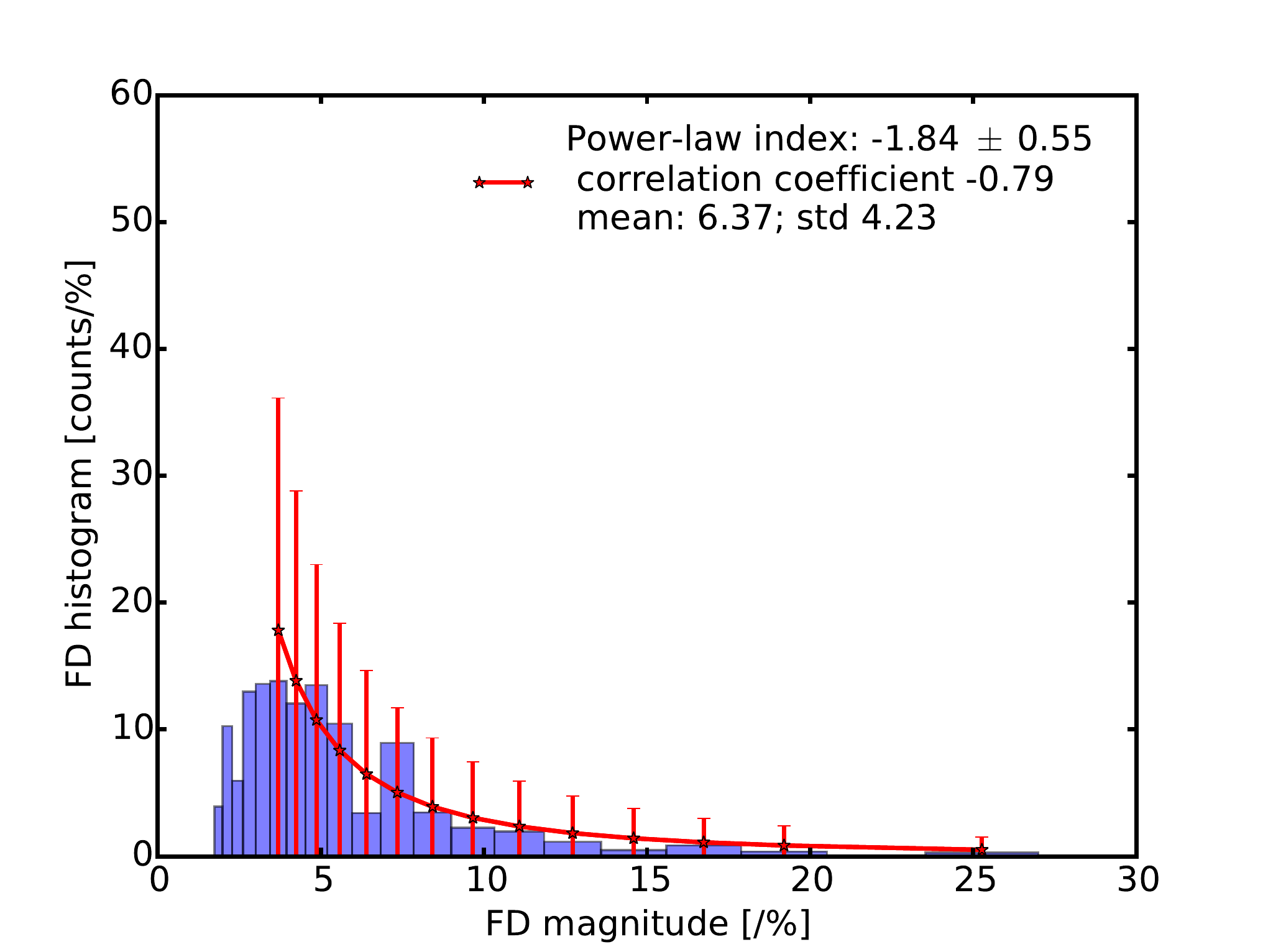}}\\
\subfloat[FD histogram of MSL FDs also seen at MAVEN] {\includegraphics[trim=5 0 20 20,clip, scale=0.45] {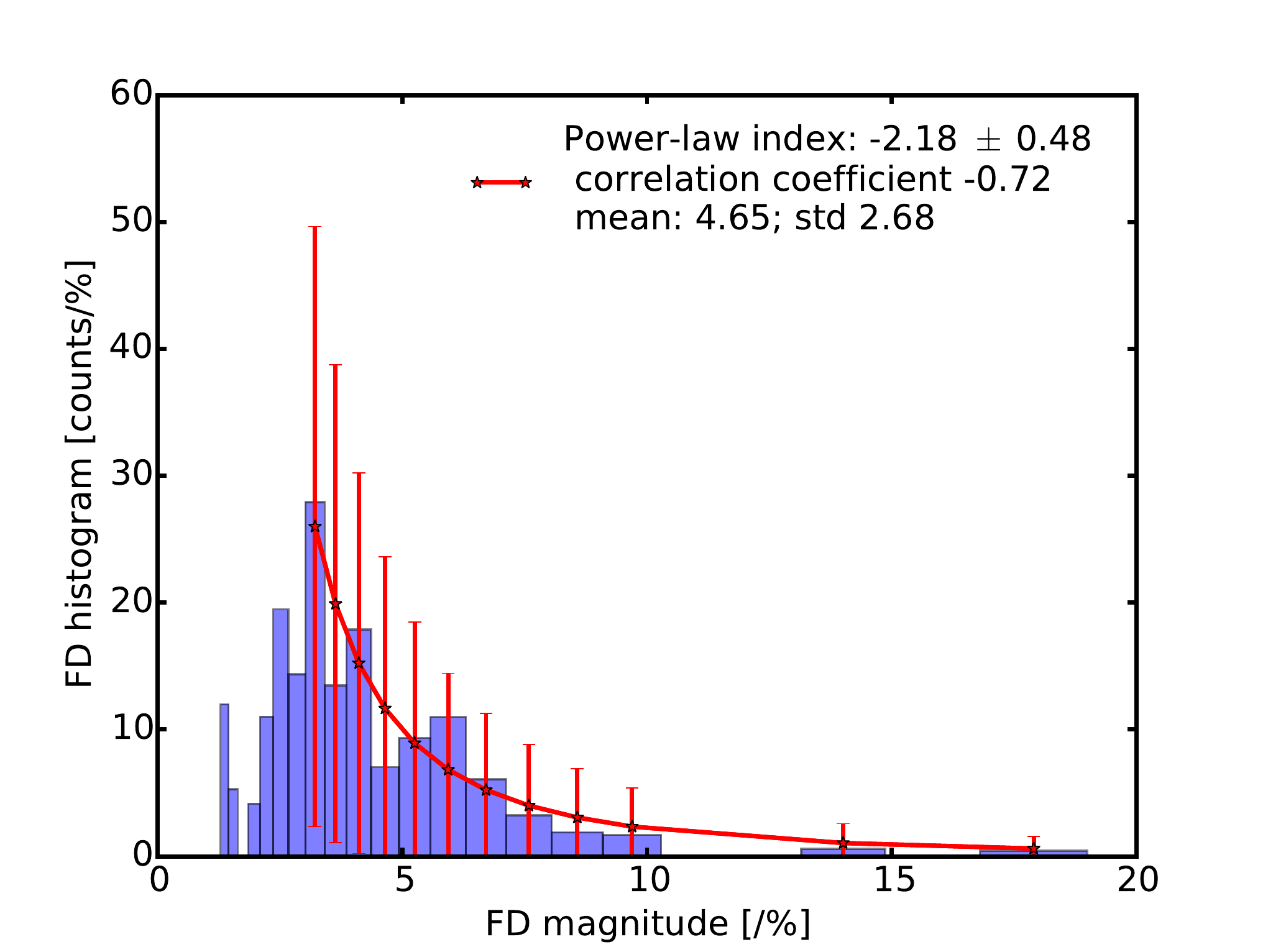}} &
\subfloat[FD histogram of MSL FDs not seen at MAVEN] {\includegraphics[trim=5 0 20 20,clip, scale=0.45] {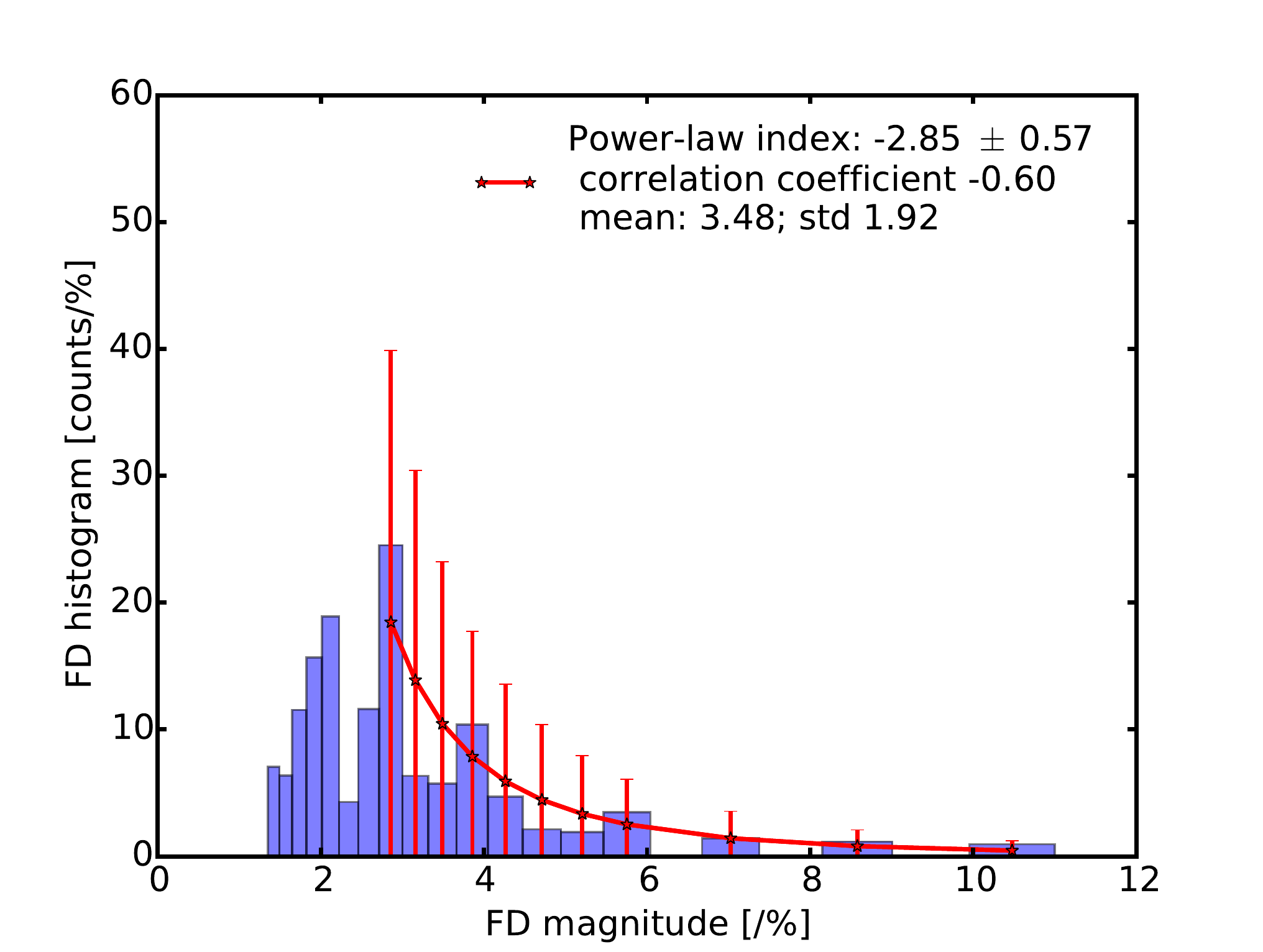}} \\
\end{tabular}
\caption{Histogram of FD magnitude at Mars (a) for all FDs seen at MSL from October 2014 until September 2016; (b) for all FDs seen at MAVEN in the same period; (c) for FDs seen at MSL which are also seen at MAVEN and (d) for FDs seen at MSL which are not seen at MAVEN. X-axes are the FD magnitude in units of drop percentage [\%] and the bins for the histogram are in logarithmic scale. Y-axes are normalized histogram [counts per \%]. 
The blue bars are the histogram while the red curves are the power-law fits and the fitting uncertainties. The legends show 1) the fitted indices and their uncertainties, 2) the correlation coefficient between the FD magnitudes and their distributions and 3) the mean value and the standard deviation of the FD distributions in each case.}\label{fig:msl_maven_fd_hist}
\end{figure*}


The final merged list contains a set of 121 FD events from October 2014 until September 2016. The histograms of their magnitudes (percentage of the maximum variation with respect to the onset value) are plotted and fitted for 4 different cases in Figure \ref{fig:msl_maven_fd_hist}: (a) for all 121 FDs seen at MSL; (b) for all 77 FDs seen at MAVEN (which is a subset of events seen at MSL); (c) for these 77 FDs at MSL and (d) for the rest FDs seen at MSL which are not seen at MAVEN excluding 5 events during MAVEN data gaps (indicated as "no data" in the list). 
The mean value and the standard deviation of the FDs in each case are noted down as legends in the figure. 
Due to much larger population of smaller events, the histograms are generated in logarithmic scales of the FD magnitudes and the y-axes are scaled with the bin-width so that they are in units of counts per drop percentage.
A power-law distribution is fitted to each histogram starting from the peak of the histogram at $\sim 3 \%$ while smaller magnitudes are not considered in the fitting to reduce potential selection biases against small events which are not much larger than the background noise. 
We have included the uncertainties of drop rates propagated through the fitting process and the resulting power-law indices; and their uncertainties are marked as legends in each panel of Figure \ref{fig:msl_maven_fd_hist}. The correlation coefficients of the FD magnitudes and their distributions are also shown in each panel. 

It is shown that the magnitude distribution of FDs at Mars, both on the surface and outside the Martian atmosphere, can be fitted with a power-law function, similar to those observed by neutron monitors at Earth \citep{belov2008forbush}.
The power-law indices for each case in Figure \ref{fig:msl_maven_fd_hist} are mostly comparable to each other within uncertainties. However, it does seem that case (b) for FDs at MAVEN has a slightly flatter distribution and MAVEN detects a smaller portion of weak FDs, likely due to the same reasons that MAVEN detects fewer FDs as explained in Section \ref{sec:selectionFDs}. Meantime MSL events not seen at MAVEN shown in case (d) have a mean magnitude slightly smaller than other cases. 
But the spectra in case (d) is less of a power-law distribution with bigger uncertainties of the fitting.  

The event on 17th October 2014 has a drop ratio of 18.10\% at MSL and 26.93 \% at MAVEN and it is the deepest FD event in our data set, associated with one of the biggest solar storms in the current solar cycle. 
In fact this event has been studied in detail by \citet{witasse2017interplanetary} using multi-spacecraft observations. The ICME was ejected at the Sun on 14 October 2014 and was detected by STEREO-A at 1 au on 16 October; it hit Mars on 17 October as observed by the Mars Express, MAVEN, Mars Odyssey and MSL missions; on 22 October it was seen by Rosetta near comet 67P/Churyumov-Gerasimenko at 3.1 au; and it was even detected by Cassini around Saturn at 9.9 au on 12 November. 

The \citet{witasse2017interplanetary} study compared the FDs detected at Mars, {comet 67P} and at Saturn: at Mars the FD seen by MSL has a magnitude of $\sim 19 \%$ (which is slightly higher than in the current list since the un-filtered data was used therein and the daily oscillation had enhanced the FD magnitude), at Rosetta the FD magnitude is about 17\% and finally at Cassini this was 15\%.
It seems to show that the FD magnitude decreases as the ICME and its shock-front propagate outwards in the heliosphere.
It is however important to note that the magnitude of RAD detected FDs on the surface of Mars may differ from that in the interplanetary space near Mars. 
This is because (1) the dose rate is not a direct measurement of particle fluxes, but a convolution of the incoming particle spectra and their energy loss inside the detectors \citep{hassler2012} and (2) more importantly, GCRs seen by MSL are modified by the Martian atmosphere and the heliospheric magnetic fields concurrently \citep{guo2015modeling}. 
The interplanetary GCR flux is first influenced by ICMEs and further modulated by the Martian atmosphere and the degree of atmospheric modulation depends on the shape of the GCR spectra on top of the atmosphere.
Specifically, GCR particles passing through the Martian atmosphere may interact with the atmospheric molecules (95\% of CO$_2$) through ionization and lose some of their energy or even stop before reaching the surface at Gale Crater such as protons below $\sim 150$ MeV; particles with high nuclear charges may also fragment into lighter ones. 
During the passage of an ICME, the original particle flux, especially of the middle-low energy part ($\le \sim$ GeV/nuc), is largely suppressed by the ICME and/or its associated shock. The following atmospheric modification of such ICME-modulated GCRs is different from that of GCRs during solar quiet time.

In order to better understand the atmospheric effects on the GCRs and to quantify the FD magnitudes in the interplanetary space from surface measurement, we analyze the 77 FDs in our list and compare the FD magnitudes detected by MSL and MAVEN for each event as shown in the next section. 


\subsection{Comparing FDs at Martian surface and those on top of the atmosphere}\label{sec:modification-atmosphere}

\begin{figure}[htb!]
\centerline{\includegraphics[width=0.55\textwidth]{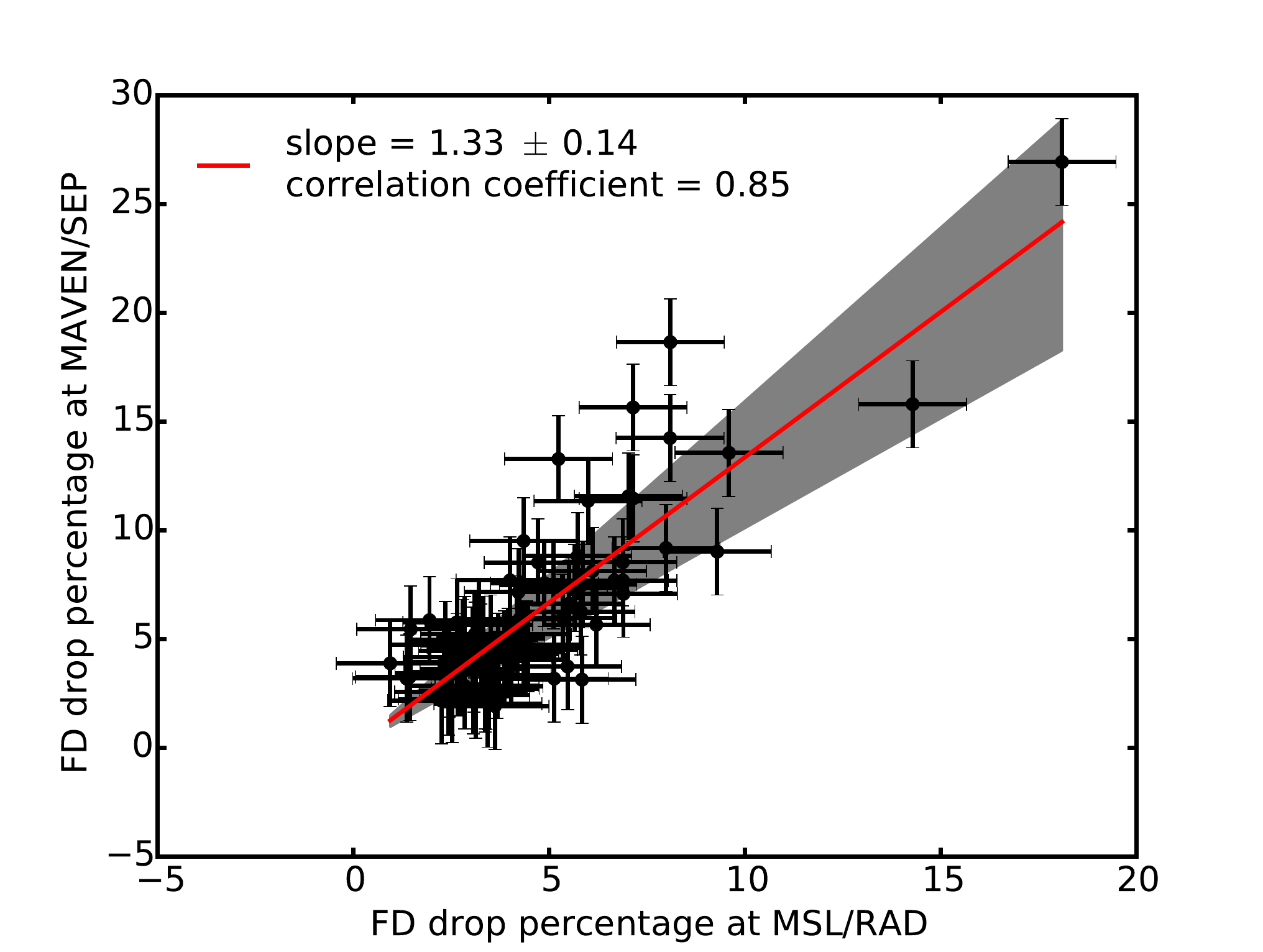}}\caption{\textit{Black:} the drop percentage for each FD detected by MSL/RAD on the Martian surface compared to that in the SEP/FTO integrated count rate of MAVEN orbiting Mars. 
\textit{Red line:} the averaged ratio (slope of a linear fit) of the FD drop percentage at MAVEN to that at MSL. \textit{Gray area:} the uncertainty range of the fitted slope obtained via a bootstrap Monte Carlo approach (more explains can be found in the text) .}\label{fig:msl_maven_fds}
\end{figure}

For each of the events seen by both MSL and MAVEN as shown in the appendix list, we plot the FD drop percentage at the FTO integrated channel (channel 5 - 10) of MAVEN/SEP versus the FD drop percentage at MSL/RAD in Figure \ref{fig:msl_maven_fds} including the error bars estimated with the selection method described in Section \ref{sec:selectionFDs}. 
The correlation coefficient between the FD magnitudes at MSL and MAVEN is as good as 0.85 and can be well fitted by a linear function (with the offset set to be zero) shown in a solid-red line in the plot. We used a bootstrap Monte Carlo approach in order to best estimate the propagated errors of the fit shown as a gray area. 
Specifically, only half of the data points are randomly selected each time and the drop percentage uncertainties shown in Table \ref{table:uncertainty} are included to generate a new sub dataset for the fitting to obtain a fitted parameter, i.e., the slope $a_1$. 
A further low-magnitude cutoff on the MSL drop percentage is applied to this subset data in order to reduce the selection biases of small events and only events with FD magnitudes larger than 5\% at MSL are used for the linear fit to obtain another slope $a_2$. 
The same cutoffs on the MAVEN drop percentage are applied to the subset data for another fitted slope $a_3$. 
The above three fittings on a random half of the dataset is repeated multiple times (200 times in this study) so that 600 fitted parameters are obtained and their average value can be taken as the final result of the slope while their standard deviation gives the uncertainty of the fit. 

The final fitted correlation between FD drop percentage at MSL and that at MAVEN is about $1.33 \pm 0.14$, i.e., an FD event seen on the surface of Mars by MSL with $\sim$ 10\% drop magnitude corresponds to $\sim$ 13.3 \% drop magnitude of GCRs with energies $\ge$ 100 MeV outside Mars' atmosphere. 
This empirical correlation is consistent with our understanding of the Marian atmospheric modulations of the incoming particles \citep{guo2017dependence}: as the atmosphere slows down particles, it 'shifts' the response weighting of the original GCRs from low-energy GCRs to middle- and high-energy GCRs; meantime the heliospheric modulation of GCRs is energy-dependent with low-energy particles more deflected and their fluxes more suppressed by the increased magnetic fields; as a result, the FDs on the surface of Mars are a better proxy of the modulation of higher-energy GCRs which are less depressed by the enhanced heliospheric activities and therefore the magnitude of the same FD on the Martian surface is smaller than that outside the Martian atmosphere. 

Such an atmospheric effect on FDs can also be readily seen by the histograms in Figure \ref{fig:msl_maven_fd_hist} (b) and (c) where the mean values of the FDs at MAVEN is 6.37\% while the same FDs have an average magnitude of 4.65 \% at MSL. This results in a ratio difference of 1.37, consistent with the fitted correlation factor of 1.33 within uncertainties.

Furthermore we want to stress again that dose rate is not a direct measurement of particle fluxes. To investigate how our resulting correlation may depend on this, we used the count rate in the same RAD dose detector for all the 121 FDs and found a correlation of 0.78 $\pm$ 0.03 for the same FD magnitudes in RAD count rate instead of RAD dose rate. 
This would result in a ratio difference of about 1.71 compared with the MAVEN seen magnitudes, further enhancing the atmospheric influence on the FDs. 
As atmospheric modification of the GCR spectra is not a simple linear process and is highly energy-dependent, the quantification of atmospheric influence on the magnitudes of FDs depends on both of the specific cutoff energies used for measurement in orbit and on ground.
The current study utilizes the ground-measured dose rate and in-orbit detected protons with 100 MeV cutoff following the conviniently accessible data products of MSL/RAD and MAVEN/SEP. 

\section{Summary and Discussion}\label{sec:discussion}

Similar to neutron monitors at Earth, RAD sees many Forbush decreases in the GCR induced surface fluxes and dose rates since the successful landing of MSL in August 2012. 
{Above} the Martian atmosphere, the {orbiting} MAVEN spacecraft has been monitoring the solar wind and IMF conditions with the SWIA, MAG and SEP instruments since September 2014. The combination of MSL/RAD and MAVEN observations gives us a great opportunity to study space weather conditions such as the passage of ICMEs, SIRs/CIRs and their associated FDs at the red planet, which is of great interest for human deep space exploration. 

For the first time, we study the statistics and properties of a list of FDs observed in-situ at Mars, both on the surface seen by MSL/RAD and at the orbit of MAVEN detected by the SEP instrument. The MSL/RAD data has a {diurnal} variation due to the Martian thermal tide and this daily oscillation has been filtered out using a notch filter tuned to remove all the multiples of 1 sol harmonics. Such filtered data are much more reliable for identifying FDs, especially the ones with smaller amplitudes.
The MAVEN/SEP FTO detected energy channels 5 to 10 are exploited for the current study as they correspond to high-energy particles which are a better proxy for GCRs. 
{Although the measured counts from SEP electrons ($\ge$ 600 keV) are not distinguished from the SEP protons ($\ge$100 MeV) in the FTO measurements (detected energy channels 5-10), as shown in the Appendix we establish that the FTO enhancements shown in this study are most likely due to SEP electron events.}

Using two years of continuous GCR measurement at Mars from September 2014 until September 2016, we have obtained the statistics for FDs both on Mars surface and outside the Martian atmosphere. 
In order to better estimate the uncertainties of our analysis, we used two independently selected FD lists and extracted the differences of the commonly selected events to generate the parametric errors of our FD selections. Such errors are applied and propagated throughout the statistic analysis of the FDs. 

The magnitudes of the set of FDs at both MSL and MAVEN can be fitted reasonably well by a power-law distribution. 
The systematic difference between the magnitudes of the same FDs within and outside the surface atmosphere may be mostly attributed to the energy-dependent modulation of the GCR particles by both the Martian atmosphere and solar activity. 
Since Mars does not have a global magnetic field like Earth and only possesses some very weak and remanent magnetic fields, the shielding of energetic GCR particles by Martian magnetic fields can be ignored. 
However the thin Martian atmosphere can still shield away some lower-energy particles (protons below $\sim$ 150 MeV) and modify the incoming GCR spectra so that particles detected by MSL/RAD on the surface of Mars are more likely generated from higher-energy GCRs which are less modulated by heliospheric magnetic fields compared to lower-energy ones. 
By comparing the magnitudes of the same FDs seen by MSL and by MAVEN as shown in Figure \ref{fig:msl_maven_fds}, we obtain an average ratio showing that an FD seen in the dose rate of MSL/RAD on the surface of Mars is about 1/1.33 times of the FD in the interplanetary space with GCR protons $\ge \sim 100$ MeV. 
Such a factor should be applied when using MSL/RAD FD data (surface data available since August 2012) to study
the arrival and impact of ICMEs or SIRs at Mars and the subsequent modulations of GCRs, especially when compared to the magnitudes of FDs caused by the same heliospheric event at other locations \citep[e.g.,][]{witasse2017interplanetary}. 

\citet{belov2008forbush} has studied a database of 5900 FD events observed at Earth (with $\sim$ 10 GV rigidity cutoffs or $\simeq 10$ GeV cutoff energies for protons) from 1957 until 2006 and systematically analyzed the distribution of the FD magnitudes which was fitted by a power law with an index of 3.1 $\pm$ 0.1. 
They show that at Earth, FDs with magnitude of larger than 3\% correspond to strong geomagnetic storms and such events occur once every 36 days on average. FDs of more than 12.5\% correspond to extreme magnetic storms which occur on average once every three years. 
In comparison, we show that FDs at Mars have a flatter power-law distribution with a bigger portion of events with larger amplitudes. The majority of the FD events on the surface of Mars have a drop percentage smaller than 10\%, except for 3 out of 121 events in two years. 
There are 79 FDs with magnitudes larger than 3\% seen by RAD and they occur once each 9 days on average, more frequent than those at Earth \footnote{Note that our data is collected close to the solar maximum (during the declining phase of the current solar cycle) where more intensive ICMEs have taken place. A comparison of FD properties and statistics during similar solar activities should be carried out in the future.}.
Given that both the ICME strength and speed should generally diminish as it propagates outwards through the heliosphere, the statistical result of FDs having larger amplitudes at the surface of Mars compared to those with 10 GV rigidity cutoffs at Earth may be largely attributed to the shielding of GCRs by the magnetosphere and much higher energies of the the primary GCRs concerned in the work by \cite{belov2008forbush}. 

Moreover, as one of the main motivations for this study, we would like to stress that when using the particle measurements and associated FDs to study the heliospheric modulation of GCRs, it is essential to compare the fluxes of GCRs with similar energy ranges at different spacecraft/locations because of spectral modulation of GCRs.


\begin{acknowledgement}
	RAD is supported by the National Aeronautics and Space Administration (NASA, HEOMD) under Jet Propulsion Laboratory (JPL) subcontract \#1273039 to Southwest Research Institute and in Germany by DLR and DLR's Space Administration grant numbers 50QM0501, 50QM1201 and 50QM1701 to the Christian Albrechts University, Kiel. 
	This work is inspired during the collaborations with Reka Winslow and Oliver Witasse using RAD Forbush decreases interpreting ICME propagations.
	We thank Bernd Heber for Patrick K\"uhl for helpful suggestions of using the EPHIN data which can be accessed here http://www2.physik.uni-kiel.de/soho/phpeph/ephin.htm.
	JG and RW acknowledge stimulating discussions with the ISSI team "Radiation Interactions at Planetary Bodies" and thank ISSI for its hospitality.
	Simulation results have been provided by the Community Coordinated Modeling Center at Goddard Space Flight Center through their archive of real-time simulations (\texttt{http://ccmc.gsfc.nasa.gov/missionsupport}). 
	The MSL and MAVEN data used in this paper are archived in the NASA Planetary Data System’s Planetary Plasma Interactions Node at the University of California, Los Angeles. 
	The PPI node is hosted at \url{https://pds-ppi.igpp.ucla.edu/}.
	The usage of the MAVEN/MAG data has been consulted with the instrument PI John Connerney.
\end{acknowledgement}

\bibliographystyle{aa}

\appendix{{Appendix: Electrons in the FTO channels of MAVEN/SEP}}\label{sec:electrons}
\begin{figure*}[htb!]
	\begin{tabular}{cc}
		\subfloat[Simulated heliospheric conditions] {\includegraphics[trim=0 0 520 0,clip, scale=0.35]{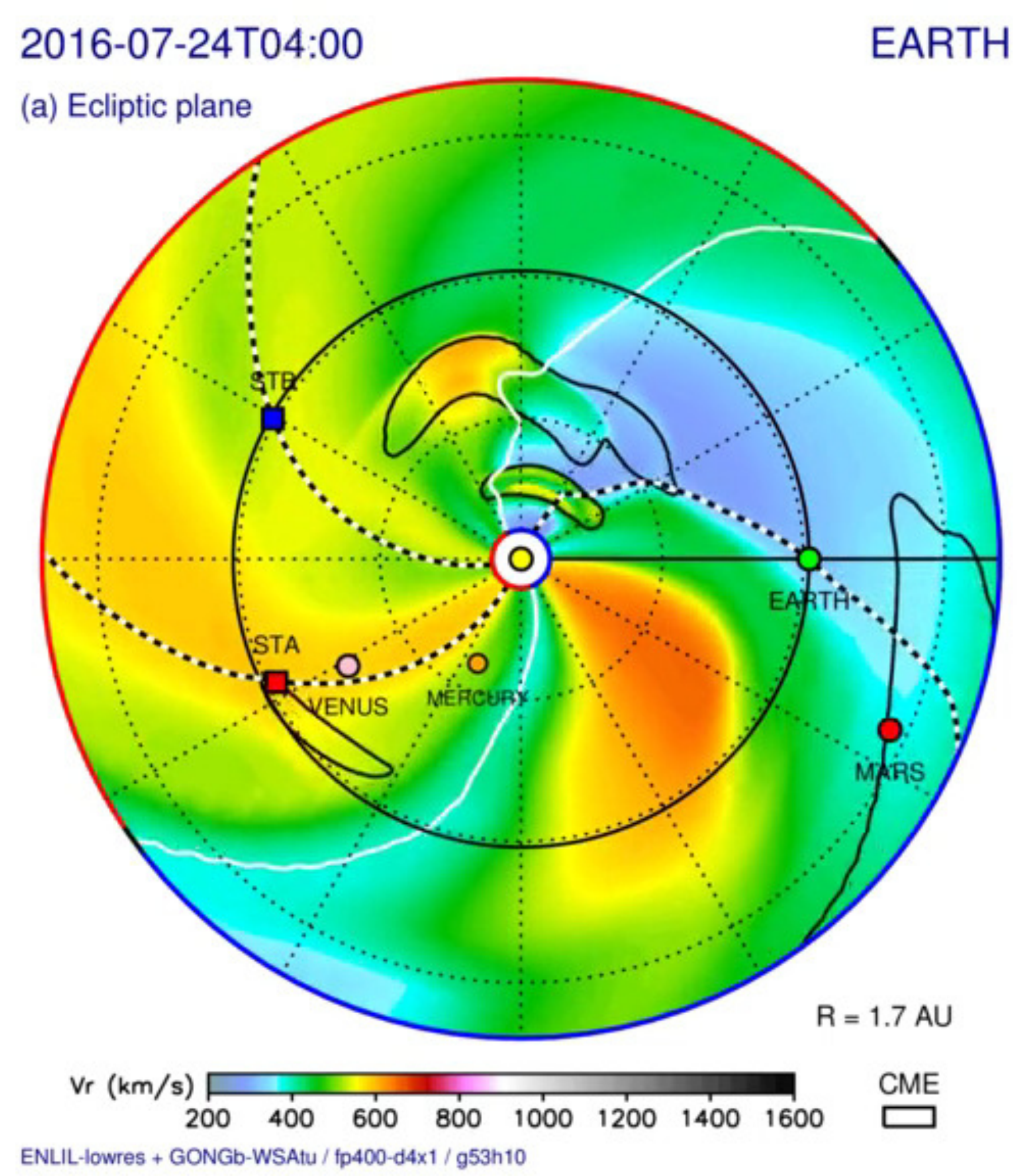}} &
		\subfloat[Comparing Earth and Mars measurement] {\includegraphics[trim=0 0 0 0,clip, scale=0.45]{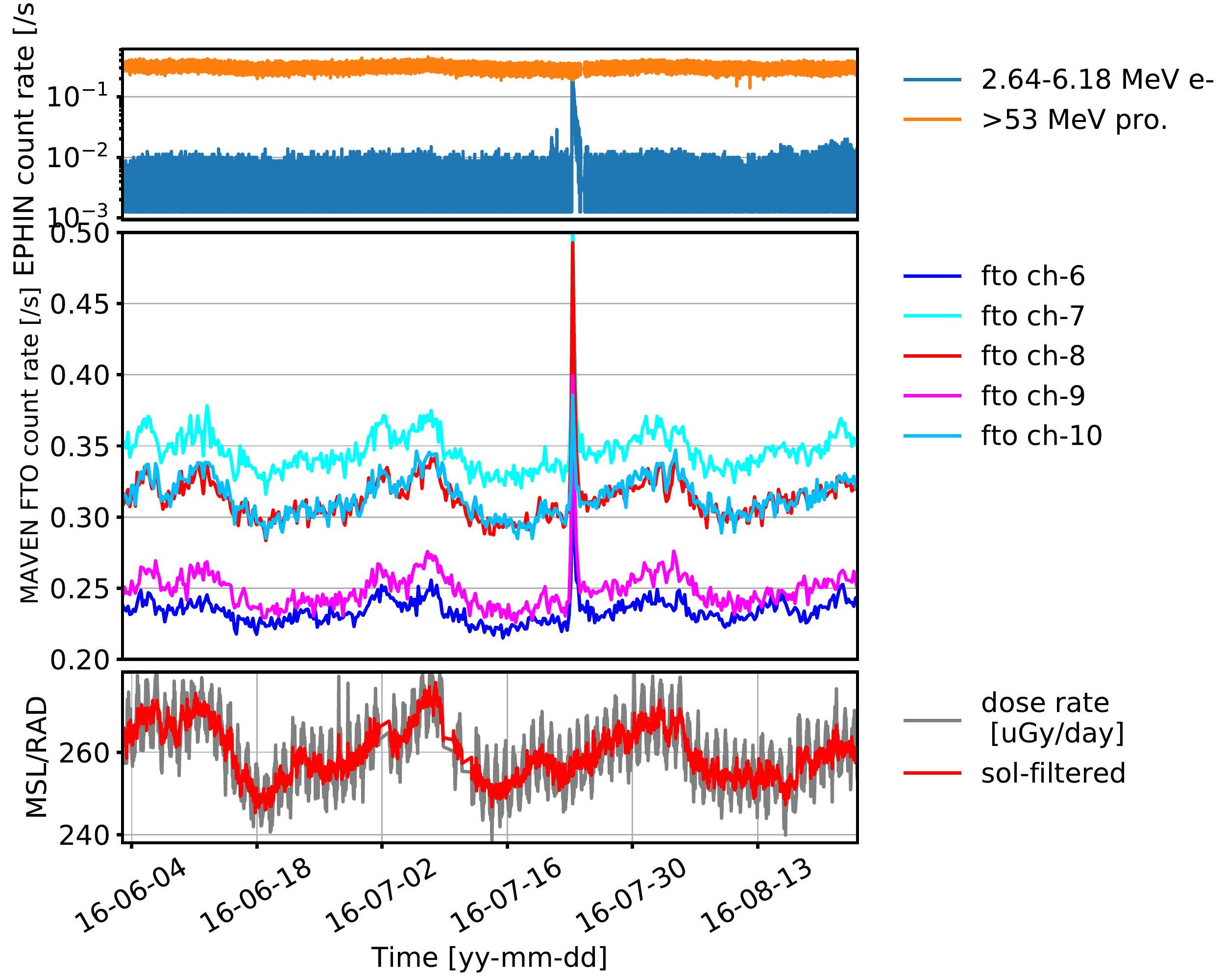}} \\
	\end{tabular}
	\caption{\textit{Left:} The WSA-ENLIL+Cone simulations of the heliospheric condition from 21.5 solar radii to 1.7 AU on 2016-07-24 (adapted from http://helioweather.net/archive/2016/07/). \textit{Right top:} The electron (blue) and proton (orange) channels of SOHO/EPHIN measurement at Earth. \textit{Right middle:} The MAVEN/SEP FTO channels from 6 to 10. \textit{Right bottom:} The MSL/RAD measurement on the surface of Mars. Detailed descriptions of the measurement can be found in the text. }\label{fig:electrons}
\end{figure*}

There were no major solar particle events (SPEs) seen by MSL/RAD on the surface of Mars during the 2-year period as shown in Fig. \ref{fig:maven-msl-all}. However there are many peaks in the MAVEN FTO integrated count rate indicating enhanced particle fluxes associated with energetic particles. As the minimum energy required for protons to penetrate through the Martian atmosphere to contribute the surface radiation environment can be as little as a hundred MeV \citep{guo2017generalized}, it is not reasonable that the MAVEN FTO channels detected these 'protons' while RAD missed all of them. 
Alternatively, electrons are likely the main contribution to these peaks as the ones with energies larger than 600 keV are also contributing to the FTO channels used here. 
During solar quiet time, this contribution is relatively small. However during a solar event (such as interplanetary shocks or SIRs) where electrons are more abundant this may lead to the peaks seen by MAVEN. 

To verify this, we use a period during July 2016 when Earth and Mars were magnetically well connected to check whether measurements carried out by the Electron Proton Helium Instrument (EPHIN) onboard the Solar and Heliospheric Observatory (SOHO) near Earth have detected electrons or protons. 
We also turn to the WSA-ENLIL+Cone model simulations as a reference for understanding possible heliospheric activities during the period. The above investigations are better demonstrated in Figure \ref{fig:electrons}. The left panel shows the WSA-ENLIL+Cone simulations of the heliospheric condition in the ecliptic plane on 24th of July 2016. 
It is shown that Earth and Mars are magnetically well connected and a strong HSS region passing both planets is indicated by the enhanced solar wind speed. This region has been corotating with the Sun for several periods and the 27-day periodic oscillations in the GCR fluxes are clearly seen by both MSL and MAVEN at Mars (panel b). 

A peak clearly present in the MAVEN FTO data around 24th of July is not visible in the RAD data. 
This peak is also present in the SOHO/EPHIN count rate of the 2.64-6.18 MeV electron channel\footnote{There are 4 different electron channels at EPHIN ranging from 0.25 to 10.4 MeV. The channel used here is the one that is least likely to be contaminated by other types of particles, such as protons}, but not in the $\ge$ 53 MeV integrated proton channel. 
As Earth and Mars are well connected, this suggests that the peaks in the MAVEN FTO channels are also caused by electrons rather than protons. 
Further investigation of the lower energy channels of the EPHIN data shows that this peak is related to a moderate solar flare also detected by the Reuven Ramaty High Energy Solar Spectroscopic Imager \footnote{https://hesperia.gsfc.nasa.gov/hessidata/dbase/hessi\_flare\_list.txt}, during which electrons upto 10 MeV and protons upto 50 MeV are accelerated. 

\onecolumn
\appendix{{Appendix: List of FD events at Mars} detected at MSL (with some also seen at MAVEN) from 2014 October until 2016 September. From left to right columns, it is shown the number of the event, the onset time of the FD, the nadir time of the FD, MSL/RAD dose rate at the onset, MSL/RAD dose rate at the nadir point, FD amplitude (drop percentage) at MSL/RAD, MAVEN/SEP integrated ($\geq 100$ MeV) count rate at the onset and at the nadir point, and finally FD amplitudes at MAVEN. }\label{sec:Fd_list}

\begin{longtable}{c|c|c|c|c|c|c|c|c}
Num & Time of Onset & Time of nadir & Dose rate & Nadir & Drop \% & flux bef-& Nadir & Drop \%\\ 
& year-mm-dd hh:mm & year-mm-dd hh:mm & ore onset & dose rate & at MSL & ore onset & flux & at MAVEN \\ 
\hline
\endhead
1 & 2014-10-17 20:45 & 2014-10-19 19:00 & 232 & 190 & 18.10 & 1.40 & 1.02 & 26.93 \\ 
2 & 2014-11-04 23:15 & 2014-11-05 18:45 & 216 & 208 & 3.70 & no FD & no FD & no FD \\ 
3 & 2014-11-08 10:15 & 2014-11-09 07:00 & 218 & 206 & 5.50 & 1.21 & 1.13 & 6.61 \\ 
4 & 2014-11-14 18:30 & 2014-11-19 13:30 & 235 & 215 & 8.51 & no data & no data & no data \\ 
5 & 2014-11-23 06:00 & 2014-11-29 14:00 & 221 & 199 & 9.95 & no data & no data & no data \\ 
6 & 2014-12-12 03:00 & 2014-12-15 11:15 & 225 & 210 & 6.67 & 1.21 & 1.12 & 7.69 \\ 
7 & 2014-12-17 12:45 & 2014-12-21 09:00 & 217 & 194 & 10.60 & increase & increase & increase \\ 
8 & 2015-01-09 03:00 & 2015-01-10 00:00 & 221 & 218 & 1.36 & 1.20 & 1.16 & 3.18 \\ 
9 & 2015-01-13 03:30 & 2015-01-15 04:00 & 218 & 203 & 6.88 & 1.26 & 1.15 & 8.52 \\ 
10 & 2015-01-18 00:45 & 2015-01-19 06:30 & 210 & 199 & 5.24 & 1.18 & 1.03 & 13.27 \\ 
11 & 2015-02-07 05:45 & 2015-02-13 12:45 & 224 & 208 & 7.14 & 1.20 & 1.06 & 11.45 \\ 
12 & 2015-02-14 04:15 & 2015-02-15 22:30 & 212 & 202 & 4.72 & 1.16 & 1.06 & 8.51 \\ 
13 & 2015-02-17 20:30 & 2015-02-18 11:00 & 205 & 202 & 1.46 & 1.12 & 1.06 & 5.44 \\ 
14 & 2015-02-21 22:00 & 2015-02-23 12:00 & 211 & 207 & 1.90 & increase & increase & increase \\ 
15 & 2015-02-27 16:00 & 2015-03-01 16:00 & 211 & 205 & 2.84 & 1.15 & 1.09 & 4.96 \\ 
16 & 2015-03-03 11:15 & 2015-03-05 13:45 & 210 & 180 & 14.29 & 1.14 & 0.96 & 15.79 \\ 
17 & 2015-03-06 06:15 & 2015-03-07 17:00 & 190 & 180 & 5.26 & increase & increase & increase \\ 
18 & 2015-03-08 10:15 & 2015-03-09 19:15 & 185 & 174 & 5.95 & increase & increase & increase \\ 
19 & 2015-03-19 12:30 & 2015-03-20 01:00 & 204 & 201 & 1.47 & no FD & no FD & no FD \\ 
20 & 2015-03-21 00:30 & 2015-03-22 06:30 & 205 & 195 & 4.88 & 1.07 & 0.99 & 7.56 \\ 
21 & 2015-03-24 22:00 & 2015-03-25 17:45 & 208 & 200 & 3.85 & increase & increase & increase \\ 
22 & 2015-03-28 10:00 & 2015-03-30 23:00 & 209 & 191 & 8.61 & increase & increase & increase \\ 
23 & 2015-04-03 02:45 & 2015-04-05 00:15 & 198 & 189 & 4.55 & no data & no data & no data \\ 
24 & 2015-04-07 03:00 & 2015-04-07 08:00 & 194 & 190 & 2.06 & no data & no data & no data \\ 
25 & 2015-04-15 20:00 & 2015-04-16 12:30 & 206 & 202 & 1.94 & 0.82 & 0.77 & 5.86 \\ 
26 & 2015-04-19 11:30 & 2015-04-21 05:30 & 209 & 206 & 1.44 & 0.86 & 0.83 & 3.24 \\ 
27 & 2015-04-23 19:15 & 2015-04-24 15:00 & 210 & 193 & 8.10 & 1.10 & 0.90 & 18.64 \\ 
28 & 2015-04-26 10:45 & 2015-04-28 05:15 & 205 & 193 & 5.85 & no FD & no FD & no FD \\ 
29 & 2015-05-02 15:00 & 2015-05-03 00:15 & 199 & 191 & 4.02 & no FD & no FD & no FD \\ 
30 & 2015-05-06 02:00 & 2015-05-06 21:45 & 194 & 186 & 4.12 & increase & increase & increase \\ 
31 & 2015-05-08 08:30 & 2015-05-08 18:30 & 194 & 190 & 2.06 & increase & increase & increase \\ 
32 & 2015-05-12 13:30 & 2015-05-13 10:00 & 203 & 197 & 2.96 & no FD & no FD & no FD \\ 
33 & 2015-05-20 09:30 & 2015-05-21 06:00 & 213 & 211 & 0.94 & 1.16 & 1.11 & 3.88 \\ 
34 & 2015-05-25 10:30 & 2015-05-25 23:00 & 216 & 210 & 2.78 & no FD & no FD & no FD \\ 
35 & 2015-06-12 07:00 & 2015-06-15 02:00 & 227 & 214 & 5.73 & 1.23 & 1.12 & 8.81 \\ 
36 & 2015-06-22 11:30 & 2015-06-23 12:15 & 224 & 208 & 7.14 & 1.28 & 1.08 & 15.64 \\ 
37 & 2015-06-28 07:30 & 2015-06-28 15:30 & 213 & 208 & 2.35 & no FD & no FD & no FD \\ 
38 & 2015-06-30 14:45 & 2015-07-01 15:00 & 214 & 200 & 6.54 & no data & no data & no data \\ 
39 & 2015-07-02 12:00 & 2015-07-03 06:30 & 208 & 201 & 3.37 & 1.14 & 1.11 & 2.72 \\ 
40 & 2015-07-08 00:30 & 2015-07-10 09:00 & 215 & 211 & 1.86 & no FD & no FD & no FD \\ 
41 & 2015-07-12 12:45 & 2015-07-13 03:45 & 215 & 209 & 2.79 & no FD & no FD & no FD \\ 
42 & 2015-07-24 06:00 & 2015-07-24 13:00 & 227 & 223 & 1.76 & no FD & no FD & no FD \\ 
43 & 2015-07-26 12:30 & 2015-07-27 11:00 & 230 & 223 & 3.04 & 1.21 & 1.16 & 4.46 \\ 
44 & 2015-08-01 17:00 & 2015-08-02 09:00 & 226 & 219 & 3.10 & 1.19 & 1.13 & 5.21 \\ 
45 & 2015-08-06 09:30 & 2015-08-09 23:00 & 225 & 219 & 2.67 & 1.24 & 1.20 & 3.46 \\ 
46 & 2015-08-16 18:00 & 2015-08-17 14:30 & 230 & 225 & 2.17 & no FD & no FD & no FD \\ 
47 & 2015-08-20 10:30 & 2015-08-26 11:30 & 230 & 217 & 5.65 & 1.21 & 1.12 & 7.34 \\ 
48 & 2015-08-28 19:45 & 2015-08-31 12:30 & 224 & 214 & 4.46 & 1.20 & 1.15 & 4.58 \\ 
49 & 2015-09-07 14:00 & 2015-09-07 19:00 & 229 & 225 & 1.75 & no FD & no FD & no FD \\ 
50 & 2015-09-09 19:00 & 2015-09-12 21:30 & 230 & 220 & 4.35 & no FD & no FD & no FD \\ 
51 & 2015-09-16 13:00 & 2015-09-17 19:45 & 229 & 222 & 3.06 & 1.22 & 1.19 & 2.63 \\ 
52 & 2015-09-18 14:45 & 2015-09-19 14:30 & 228 & 220 & 3.51 & 1.21 & 1.15 & 5.02 \\ 
53 & 2015-10-05 09:00 & 2015-10-09 03:30 & 230 & 220 & 4.35 & 1.28 & 1.16 & 9.50 \\ 
54 & 2015-10-09 15:45 & 2015-10-11 13:00 & 225 & 216 & 4.00 & 1.25 & 1.15 & 7.69 \\ 
55 & 2015-10-18 03:00 & 2015-10-20 03:30 & 234 & 222 & 5.13 & 1.23 & 1.19 & 3.17 \\ 
56 & 2015-10-23 11:30 & 2015-10-24 10:00 & 229 & 224 & 2.18 & no FD & no FD & no FD \\ 
57 & 2015-10-29 09:30 & 2015-10-30 20:30 & 234 & 226 & 3.42 & increase & increase & increase \\ 
58 & 2015-11-08 00:30 & 2015-11-08 20:00 & 230 & 219 & 4.78 & increase & increase & increase \\ 
59 & 2015-11-13 06:45 & 2015-11-14 22:00 & 245 & 236 & 3.67 & 1.31 & 1.27 & 3.35 \\ 
60 & 2015-11-19 22:15 & 2015-11-22 06:45 & 242 & 225 & 7.02 & 1.32 & 1.17 & 11.56 \\ 
61 & 2015-11-24 05:15 & 2015-11-25 05:45 & 233 & 225 & 3.43 & 1.24 & 1.22 & 2.01 \\ 
62 & 2015-11-28 01:00 & 2015-11-28 18:30 & 238 & 232 & 2.52 & 1.25 & 1.23 & 2.23 \\ 
63 & 2015-12-01 15:00 & 2015-12-02 14:00 & 243 & 237 & 2.47 & no FD & no FD & no FD \\ 
64 & 2015-12-07 09:30 & 2015-12-08 08:15 & 249 & 240 & 3.61 & 1.36 & 1.34 & 1.91 \\ 
65 & 2015-12-12 21:00 & 2015-12-14 07:00 & 256 & 248 & 3.12 & no FD & no FD & no FD \\ 
66 & 2015-12-14 20:15 & 2015-12-15 14:45 & 253 & 242 & 4.35 & 1.42 & 1.36 & 4.50 \\ 
67 & 2015-12-15 21:00 & 2015-12-17 07:00 & 247 & 230 & 6.88 & 1.37 & 1.26 & 7.69 \\ 
68 & 2015-12-18 15:45 & 2015-12-21 23:30 & 236 & 226 & 4.24 & 1.34 & 1.26 & 5.91 \\ 
69 & 2016-01-02 16:15 & 2016-01-03 14:45 & 239 & 229 & 4.18 & 1.46 & 1.38 & 5.22 \\ 
70 & 2016-01-04 03:30 & 2016-01-04 14:30 & 238 & 230 & 3.36 & 1.40 & 1.36 & 2.86 \\ 
71 & 2016-01-05 06:45 & 2016-01-06 10:45 & 235 & 223 & 5.11 & 1.34 & 1.24 & 7.46 \\ 
72 & 2016-01-12 19:30 & 2016-01-13 04:00 & 247 & 237 & 4.05 & no FD & no FD & no FD \\ 
73 & 2016-01-16 01:30 & 2016-01-17 12:45 & 247 & 240 & 2.83 & 1.37 & 1.33 & 2.85 \\ 
74 & 2016-01-17 21:45 & 2016-01-18 09:30 & 248 & 243 & 2.02 & no FD & no FD & no FD \\ 
75 & 2016-01-20 04:15 & 2016-01-22 05:45 & 246 & 240 & 2.44 & 1.35 & 1.30 & 3.41 \\ 
76 & 2016-01-25 11:15 & 2016-01-26 19:45 & 250 & 242 & 3.20 & 1.40 & 1.32 & 5.64 \\ 
77 & 2016-01-27 20:30 & 2016-01-28 19:00 & 248 & 241 & 2.82 & no FD & no FD & no FD \\ 
78 & 2016-02-01 08:15 & 2016-02-02 09:00 & 248 & 237 & 4.44 & 1.39 & 1.32 & 4.74 \\ 
79 & 2016-02-04 00:15 & 2016-02-04 21:45 & 257 & 242 & 5.84 & 1.41 & 1.36 & 3.12 \\ 
80 & 2016-02-07 15:30 & 2016-02-09 02:15 & 253 & 243 & 3.95 & 1.43 & 1.37 & 4.40 \\ 
81 & 2016-02-11 15:45 & 2016-02-12 13:30 & 258 & 250 & 3.10 & 1.41 & 1.34 & 4.69 \\ 
82 & 2016-02-15 03:00 & 2016-02-15 13:30 & 256 & 249 & 2.73 & no FD & no FD & no FD \\ 
83 & 2016-02-17 08:30 & 2016-02-18 14:30 & 256 & 249 & 2.73 & 1.39 & 1.35 & 3.44 \\ 
84 & 2016-02-19 21:15 & 2016-02-23 09:30 & 256 & 250 & 2.34 & 1.42 & 1.35 & 4.73 \\ 
85 & 2016-02-29 08:15 & 2016-03-05 06:30 & 262 & 246 & 6.11 & 1.47 & 1.35 & 8.12 \\ 
86 & 2016-03-05 23:45 & 2016-03-06 11:15 & 252 & 244 & 3.17 & 1.37 & 1.37 & 1.374 \\ 
87 & 2016-03-10 07:45 & 2016-03-12 07:00 & 258 & 242 & 6.20 & 1.45 & 1.37 & 5.65 \\ 
88 & 2016-03-16 08:30 & 2016-03-17 08:15 & 256 & 248 & 3.12 & 1.44 & 1.41 & 2.43 \\ 
89 & 2016-03-19 14:45 & 2016-03-20 06:15 & 251 & 244 & 2.79 & 1.43 & 1.36 & 4.82 \\ 
90 & 2016-03-26 21:30 & 2016-03-28 08:30 & 258 & 243 & 5.81 & 1.44 & 1.35 & 6.25 \\ 
91 & 2016-03-28 15:30 & 2016-03-29 02:45 & 247 & 241 & 2.43 & 1.36 & 1.33 & 2.56 \\ 
92 & 2016-04-06 05:00 & 2016-04-06 17:30 & 258 & 253 & 1.94 & no FD & no FD & no FD \\ 
93 & 2016-04-13 02:30 & 2016-04-13 22:00 & 261 & 250 & 4.21 & 1.51 & 1.40 & 7.15 \\ 
94 & 2016-04-14 22:30 & 2016-04-16 19:45 & 256 & 242 & 5.47 & 1.45 & 1.39 & 3.73 \\ 
95 & 2016-04-22 10:30 & 2016-04-25 11:30 & 261 & 243 & 6.90 & 1.47 & 1.37 & 7.07 \\ 
96 & 2016-05-03 17:45 & 2016-05-05 18:00 & 259 & 249 & 3.86 & 1.47 & 1.41 & 4.29 \\ 
97 & 2016-05-09 17:30 & 2016-05-10 18:15 & 262 & 248 & 5.34 & 1.47 & 1.39 & 5.97 \\ 
98 & 2016-05-15 03:00 & 2016-05-16 08:45 & 262 & 255 & 2.67 & increase & increase & increase \\ 
99 & 2016-05-18 00:00 & 2016-05-22 23:00 & 263 & 242 & 7.98 & 1.50 & 1.37 & 9.18 \\ 
100 & 2016-05-25 07:30 & 2016-05-26 15:30 & 257 & 253 & 1.56 & no FD & no FD & no FD \\ 
101 & 2016-05-31 04:30 & 2016-06-01 05:00 & 260 & 254 & 2.31 & 1.47 & 1.41 & 3.88 \\ 
102 & 2016-06-06 19:30 & 2016-06-07 09:00 & 271 & 262 & 3.32 & 1.55 & 1.47 & 4.96 \\ 
103 & 2016-06-08 01:15 & 2016-06-08 23:00 & 268 & 261 & 2.61 & no FD & no FD & no FD \\ 
104 & 2016-06-11 03:15 & 2016-06-17 19:45 & 271 & 245 & 9.59 & 1.58 & 1.36 & 13.55 \\ 
105 & 2016-06-22 20:30 & 2016-06-25 09:15 & 260 & 252 & 3.08 & 1.46 & 1.41 & 3.62 \\ 
106 & 2016-07-03 09:45 & 2016-07-03 18:00 & 265 & 258 & 2.64 & 1.58 & 1.49 & 5.76 \\ 
107 & 2016-07-08 05:30 & 2016-07-15 22:30 & 272 & 250 & 8.09 & 1.64 & 1.40 & 14.23 \\ 
108 & 2016-07-19 21:00 & 2016-07-21 19:30 & 260 & 251 & 3.46 & 1.46 & 1.42 & 2.82 \\ 
109 & 2016-07-28 11:00 & 2016-07-29 05:15 & 266 & 260 & 2.26 & 1.52 & 1.49 & 2.17 \\ 
110 & 2016-08-02 01:30 & 2016-08-03 01:30 & 271 & 261 & 3.69 & 1.58 & 1.52 & 4.17 \\ 
111 & 2016-08-04 07:15 & 2016-08-10 11:00 & 267 & 251 & 5.99 & 1.57 & 1.39 & 11.34 \\ 
112 & 2016-08-11 22:00 & 2016-08-12 13:30 & 256 & 249 & 2.73 & no FD & no FD & no FD \\ 
113 & 2016-08-14 07:45 & 2016-08-16 02:45 & 256 & 247 & 3.52 & 1.50 & 1.45 & 1.45 \\ 
114 & 2016-08-18 11:15 & 2016-08-18 22:30 & 261 & 254 & 2.68 & 1.50 & 1.44 & 3.93 \\ 
115 & 2016-08-21 19:15 & 2016-08-23 22:30 & 264 & 257 & 2.65 & 1.56 & 1.50 & 4.16 \\ 
116 & 2016-08-29 18:30 & 2016-09-02 18:00 & 269 & 244 & 9.29 & 1.60 & 1.45 & 9.01 \\ 
117 & 2016-09-08 20:00 & 2016-09-09 06:15 & 256 & 238 & 7.03 & no FD & no FD & no FD \\ 
118 & 2016-09-16 04:45 & 2016-09-17 00:15 & 248 & 241 & 2.82 & no FD & no FD & no FD \\ 
119 & 2016-09-17 12:30 & 2016-09-18 11:00 & 248 & 238 & 4.03 & 1.58 & 1.52 & 4.04 \\ 
120 & 2016-09-22 15:45 & 2016-09-23 16:30 & 276 & 267 & 3.26 & 1.63 & 1.55 & 4.60 \\ 
121 & 2016-09-24 04:00 & 2016-09-28 01:15 & 273 & 257 & 5.86 & 1.60 & 1.48 & 7.50 
\end{longtable}

\end{document}